\documentclass[aps,prx,twocolumn,a4,showpacs,superscriptaddress]{revtex4-1} 
\usepackage{amsmath,amssymb,amsfonts,upgreek}

\usepackage[english]{babel}
\usepackage[utf8x]{inputenc}
\usepackage[T1]{fontenc}
\usepackage{bm}

\usepackage{rotating}
\usepackage[colorinlistoftodos]{todonotes}
\usepackage[colorlinks=true, allcolors=blue]{hyperref}

\usepackage{amsmath,amssymb,amsfonts,upgreek}
\usepackage{gensymb}
\usepackage{color}
\usepackage{graphicx}
\usepackage{setspace}
\usepackage{multirow}
\usepackage{hhline}
\usepackage{bm}
\usepackage{comment}
\usepackage{natbib}

\begin{document}

\title{Four-spin ring interaction as a source of unconventional magnetic orders in orthorhombic perovskite manganites}

\author{Natalya S.\ Fedorova}
\email{natalya.fedorova@mat.ethz.ch}
\affiliation{Materials Theory, ETH Z\"{u}rich, Wolfgang-Pauli-Strasse 27, CH-8093 Z\"{u}rich, Switzerland}
\author{Amad\'{e} Bortis}
\affiliation{Laboratory for Multifunctional Ferroic Materials, ETH Z\"{u}rich, Vladimir-Prelog-Weg 4, CH-8093 Z\"{u}rich, Switzerland},
\author{Christoph Findler}
\affiliation{Materials Theory, ETH Z\"{u}rich, Wolfgang-Pauli-Strasse 27, CH-8093 Z\"{u}rich, Switzerland}
\author{Nicola A.\ Spaldin}
\email{nicola.spaldin@mat.ethz.ch}
\affiliation{Materials Theory, ETH Z\"{u}rich, Wolfgang-Pauli-Strasse 27, CH-8093 Z\"{u}rich, Switzerland}
\begin{abstract}

We use \textit{ab initio} electronic structure calculations in combination with Monte Carlo simulations to investigate the magnetic and ferroelectric properties of bulk orthorhombic HoMnO$_3$ and ErMnO$_3$. Our goals are to explain the inconsistencies in the measured magnetic properties of the orthorhombic perovskite manganites (o-$R$MnO$_3$) with small rare-earth ($R$) cations or Y, as well as the contradictions between the directions and amplitudes of the electric polarizations reported by different experimental groups. Our computations stabilize several exotic magnetic orders (so-called w-spiral, H-AFM and I-AFM), whose presence resolve the contradictions in the measured magnetic and ferroelectric properties of o-$R$MnO$_3$. We show that these orders emerge due to strong four-spin ring exchange interactions.

\end{abstract}

\maketitle

\section{Introduction}
Magnetoelectric multiferroics, materials which possess magnetic and ferroelectric orders in a single phase, are the focus of intensive investigation as the coexistence and coupling between these two orders may open new avenues for the development of multifunctional devices \cite{spaldin2005multiferroics,cheong2007multiferroics,fiebig2016multiferroics,spaldin2010MF_past_present}. Those compounds in which an electric polarization is induced by an inversion-symmetry-breaking magnetic order are especially interesting as they
provide high tunability of their ferroelectric properties by applying a magnetic field or vice versa \cite{tokura2014multiferroics}. The paradigmatic representatives of this class of materials are the orthorhombic manganites, o-$R$MnO$_3$, where $R$ is a rare-earth cation, typically with relatively small radius, or Y. In o-$R$MnO$_3$ the complex interplay between lattice, spin and orbital degrees of freedom leads to the establishment of frustrated magnetic orders, such as an incommensurate (IC) spiral \cite{kimura2003tbmno3,kenzelmann2005tbmno3} or an E-AFM order \cite{munoz2001homno3}, which induce a spontaneous electric polarization $P$. The appearance of $P$ in systems with IC spiral order is usually considered to be an effect due to spin-orbit coupling \cite{katsuraPRLknb_model,sergienko2006DMI}. Since this coupling is weak, the resulting electric polarization is relatively small. For example, in TbMnO$_3$ the measured electric polarization reaches a maximum $P$$\approx$0.08 $\mu$C/cm$^2$ \cite{kimura2003tbmno3}, which is three orders of magnitude weaker than that of proper ferroelectrics.
On the other hand, it was theoretically predicted that in systems possessing E-AFM order, $P$ emerges due to symmetric exchange striction and this mechanism should provide at least two orders of magnitude larger polarization values compared to those of systems with a spiral order \cite{sergienko2006exch_strict,picozzi2007homno3}.
Following these predictions, numerous experimental studies of the magnetic and ferroelectric properties were performed for o-$R$MnO$_3$ with small $R$ cations ($R$=Ho...Lu), for which the E-AFM order was expected to be the magnetic ground state \cite{ye2007ermno3,lee2011homno3,lorenz2007homno3_ymno3}. However, as we will describe in detail in Sec.\ \ref{sec:magn_FE_rmno3}, these studies gave contradictory results for the measured values of their magnetic and ferroelectric properties. In particular, there is still no agreement on the type of magnetic ordering in o-HoMnO$_3$, o-ErMnO$_3$
 and o-YMnO$_3$, which are on the borderline between the spiral and the E-AFM phases in the magnetic phase diagram of the o-$R$MnO$_3$ series \cite{zhou2006rmno3,ishiwata2010rmno3}. Moreover, the theoretically predicted polarization values have not been experimentally observed for bulk samples of these materials and this has not been explained. There are also contradictions between the amplitudes and directions of the electric polarizations reported for these systems by different groups \cite{lorenz2007homno3_ymno3,lee2011homno3,ye2007ermno3}.
 
In this work we combine \textit{ab initio} electronic structure calculations and Monte Carlo (MC) simulations to investigate the magnetic and ferroelectric properties of bulk o-HoMnO$_3$ and o-ErMnO$_3$. We consider only effects due to the ordering of Mn$^{3+}$ spins and do not take into account those arising from the ordering of $R^{3+}$ moments. We describe the magnetism in terms of a model Hamiltonian which includes isotropic Heisenberg, biquadratic and four-spin ring exchange interactions as well as the Dzyaloshinskii-Moriya interaction (DMI) and single ion anisotropy (SIA). We extract the exchange couplings and anisotropies by mapping the results of density functional theory (DFT) calculations onto the considered model Hamiltonian and use them in a series of MC simulations in order to determine the magnetic ground states in these systems. We report several exotic magnetic orders that have not been previously identified and which are favored by strong four-spin ring exchange interaction. We show that the presence of these magnetic orders can resolve the inconsistencies in previous theoretical and experimental studies of the magnetic and ferroelectric properties of these materials.

This article is structured as follows: in Sec.\ \ref{sec:magn_FE_rmno3} we describe the crystal structure and magnetic properties of o-$R$MnO$_3$, and the possible mechanisms by which the different magnetic orders can induce electric polarization. We also summarize the existing experimental and theoretical results, which motivated this study. In Sec.\ \ref{sec:hamiltonian} we introduce the model Hamiltonian that is used to describe the magnetism in these materials. In Sec.\ \ref{sec:comput_details} we specify the computational details. In Sec.\ \ref{sec:results} we describe the analysis performed and the results obtained for o-HoMnO$_3$ and o-ErMnO$_3$. Here we also introduce three exotic magnetic orders which may be stabilized in these systems by strong four-spin ring exchange interactions. In Sec.\ \ref{sec:guide} we present the values of different observables which may help to identify these exotic magnetic orders experimentally. Finally, in Sec.\   \ref{sec:conclusions} we summarize all the key findings of our investigation.   

\begin{figure}
\includegraphics[width=0.5\textwidth]{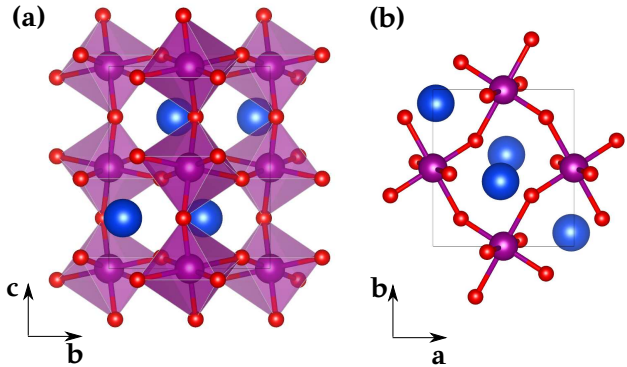}
\caption{\label{fig:cryst_struct} Crystal structure of o-$R$MnO$_3$ (\textit{Pbnm} notation): (a) view in the $bc$ plane, (b) - in the $ab$ plane. Blue spheres indicate $R$ ions, purple - Mn ions, red - O ions.}
\end{figure}

\section{Motivation and background}
\subsection{Magnetism and ferroelectricity in o-$R$MnO$_3$}
\label{sec:magn_FE_rmno3}
The orthorhombic rare-earth manganites adopt the orthorhombically distorted perovskite structure with \textit{Pbnm} (\#62) space group (see Fig.\ \ref{fig:cryst_struct}) \cite{alonso2000rmno3}. Two  primary structural distortions reduce the crystal symmetry from cubic to orthorhombic: A Jahn-Teller (JT) distortion of the MnO$_6$ octahedra \cite{kanamori1960JT} and a GdFeO$_3$-type (GFO) cooperative rotation and tilting of these octahedra \cite{woodward1997tiltings}. The strengths of these distortions across the o-$R$MnO$_3$ series are responsible for the trends in the magnetic properties. 
In o-$R$MnO$_3$ each Mn$^{3+}$ ion has four electrons in the $3d$ levels  ($t_{2g}^3e_g^1$). The cooperative JT distortion of the MnO$_6$ octahedra results in the establishment of  long-range ordering of the $e_g$ orbitals (staggered orbitals with $3x^2-r^2$/$3y^2-r^2$ character) within the $ab$ planes. According to the Goodenough-Kanamori-Anderson rules, this ordering favors ferromagnetic (FM) superexchange interactions (through the $p$ states of O$^{2-}$) between the $e_g$ spins on the nearest neighboring (NN) Mn$^{3+}$ sites within the $ab$ planes and antiferromagnetic (AFM) interactions along the $c$ axis \cite{khomskii1973orbital_ordering,goodenough1955superexchange,kanamori1959superexchange,anderson1959superexchange}. Thus, it leads to the establishment of A-AFM order \cite{Wollan1955ABC} in o-$R$MnO$_3$ with $R$=La,...,Gd. Further decrease in the radius of the $R$ cation, however, leads to a change in the magnetic ground state in o-$R$MnO$_3$. Indeed, in TbMnO$_3$ and DyMnO$_3$ an IC spiral order is stabilized at low temperatures, while in o-$R$MnO$_3$ with $R$=Ho,...,Lu early magnetic measurements reported the presence of E-AFM order \cite{Wollan1955ABC,munoz2001homno3}. Such an evolution of the magnetic order 
occurs due to the increasing GFO distortion, which is favored by the small size of the $R$ cation and results in the reduction of the Mn-O-Mn bond angles \cite{kimura2003rmno3}. This weakens the overlap between the $d$ states of Mn and $p$ states of O, which in turn decreases the strength of the FM exchange between NN Mn spins within the $ab$ planes and makes the effect of other couplings (such as further-neighbor interactions, higher-order exchanges and anisotropic coupling terms) more pronounced \cite{zhou2006rmno3}. 

In o-$R$MnO$_3$ with $R$=Tb,...,Lu the establishment of the IC spiral or E-AFM orders is accompanied by the appearance of a spontaneous electric polarization \cite{kimura2003tbmno3,lee2011homno3,ye2007ermno3,ishiwata2010rmno3}. In the systems with spiral order, emergence of the ferroelectricity is usually explained as an effect due to spin-orbit coupling. $P$ can be of purely electronic origin \cite{katsuraPRLknb_model} and can also have a contribution originating from the antisymmetric exchange striction \cite{sergienko2006DMI} (or inverse DMI), that is the displacements of the O$^{2-}$ anions such as to minimize the energy of the DMI between the spins on the neighboring magnetic sites \cite{tokura2014multiferroics}. The electric polarization created according to this mechanism can be written as follows: 
\begin{equation}
\label{eq:antisym_exch_strict}
\mathbf{P^{AS}}\propto\sum_{\left\langle i,j \right\rangle}\mathbf{e}_{ij}\times\left[\mathbf{S}_i\times\mathbf{S}_j\right],
\end{equation}
where the summation is over pairs of NN spins $\mathbf{S}_i$ and $\mathbf{S}_j$ on sites $i$ and $j$ and $\mathbf{e}_{ij}$ is the unit vector connecting sites $i$ and $j$. Based on this formula one would expect the electric polarization in a system with spiral order to be perpendicular to the propagation vector of the spiral ($\mathbf{q}$) and to the spin rotation axis. For example, for TbMnO$_3$, which possesses an IC spiral order with a propagation vector $\mathbf{q}$=(0,0.28,0) and the spins rotating within the $bc$ plane, Eq. \ref{eq:antisym_exch_strict} gives the electric polarization along the $c$ axis, which was indeed observed experimentally \cite{kimura2003tbmno3}. For an $ab$ spiral, in turn, the polarization is expected along the $a$ axis. The amplitudes of  polarizations induced by this mechanism, however, are small (three orders of magnitude smaller than those of conventional ferroelectrics), because the spin-orbit coupling, which drives them, is intrinsically weak. On the other hand, for systems with E-AFM order, an alternative mechanism inducing the electric polarization was proposed by Sergienko \textit{et al} \cite{sergienko2006exch_strict}. It is based on symmetric exchange striction leading to an increase in Mn-O-Mn bond angles between neighboring Mn ions with parallel spins and to a decrease in these angles between the ions with antiparallel spins within the $ab$ planes to minimize the energy of the Heisenberg exchange interactions. The expression for the electric polarization induced by this mechanism is:
\begin{equation}
\label{eq:sym_exch_strict}
\mathbf{P^S}\propto\sum_{\left\langle i,j \right\rangle}\mathbf{\Pi}_{ij}\left(\mathbf{S}_i\cdot\mathbf{S}_j\right),
\end{equation}
where $\mathbf{\Pi}_{ij}$ is a unit vector along one of the crystallographic directions. For E-AFM order the resulting $\mathbf{P^S}$ is parallel to the $a$ axis.
As the energy scale of the Heisenberg interactions is usually higher than that of the DMI, the amplitude of the electric polarization generated by E-AFM order is expected to be larger than that of spiral order. For example, the amplitude of $P$ predicted in Ref. \onlinecite{sergienko2006exch_strict}
for o-HoMnO$_3$ ranged between 0.5-12 $\mu$C/cm$^2$, which is at least one order of magnitude larger than the polarization measured in TbMnO$_3$.  
Later this prediction was confirmed by Berry phase calculations which gave P$\approx$6 $\mu$C/cm$^2$ for o-HoMnO$_3$ \cite{picozzi2007homno3}.

The theoretical prediction that large $P$ values should be induced by E-AFM order triggered multiple studies of the magnetic and ferroelectric properties of o-$R$MnO$_3$ with small $R$ (Ho,...,Lu and Y) for which E-AFM is expected to be the magnetic ground state \cite{lorenz2007homno3_ymno3,lee2011homno3,ye2007ermno3,ishiwata2010rmno3}. These studies, however, gave contradictory results. Indeed, there is still no agreement on the type of magnetic order in these compounds, as different magnetic structures were reported by several groups even for systems with the same $R$. For example, in early neutron diffraction measurements on powder o-HoMnO$_3$, commensurate E-AFM order (with $q_b$=0.5) of Mn$^{3+}$ spins was observed \cite{munoz2001homno3}. However, in neutron diffraction experiments performed by different groups an IC magnetic order with $q_b$$\approx$0.4 was found in this material \cite{brinks2001homno3,lee2011homno3}, and it was identified as a sinusoidal spin density wave with Mn spins aligned along the $b$ direction. Magnetic states with similar IC modulation vectors were also reported for o-YMnO$_3$ \cite{munoz2002ymno3} and o-ErMnO$_3$ \cite{ye2007ermno3}. In the latter case, however, the authors did not make a definitive conclusion about the type of the observed magnetic order. A theoretical study based on MC simulations suggested that the observed state could consist of coexisting spiral and E-AFM orders, however, this phase coexistence was metastable in these simulations \cite{mochizuki2011theory,mochizuki2010magnetostriction}. The results of measurements of the electric polarization ($\mathbf{P}$) in these materials are even more puzzling. For example, Lorenz \textit{et al.} \cite{lorenz2007homno3_ymno3} observed $P||a$ reaching a maximal value of $\approx$0.01 $\mu$C/cm$^2$ for o-HoMnO$_3$. Later Feng \textit{et al.} \cite{feng2010homno3} reported $P$ in the range of 0.01-0.07 $\mu$C/cm$^2$ (different values for differently synthesized samples) for the same material. All the measured values are much smaller than those predicted theoretically \cite{sergienko2006exch_strict,picozzi2007homno3} and this disagreement between theory and experiment is still not understood.
\begin{figure}[t]
\includegraphics[width=0.42\textwidth]{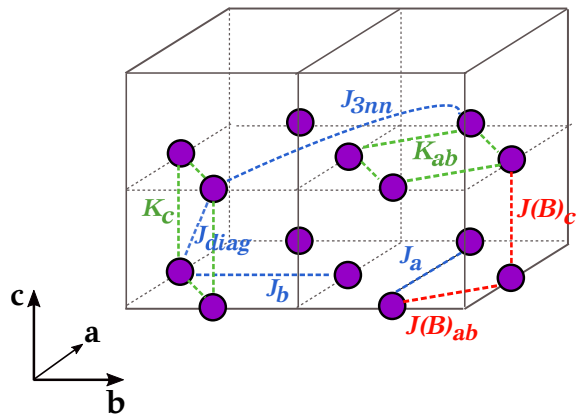}
\caption{\label{fig:exchanges} Heisenberg, biquadratic and four-spin ring exchange interactions considered in this work. A 1$\times$2$\times$1 supercell of the conventional o-$R$MnO$_3$ unit cell (only Mn ions) is shown. NN Heisenberg and biquadratic couplings are highlighted in red, NNN Heisenberg couplings in blue and four-spin ring exchanges in green.}
\end{figure}
Furthermore, Ref. \onlinecite{lorenz2007homno3_ymno3} showed an interesting temperature dependence of $P$ in o-HoMnO$_3$  - it started to increase below the lock-in temperature of the Mn$^{3+}$ spins and a sharp increase in $P$ occurred only below the ordering temperature of Ho$^{3+}$ moments, pointing to a significant role of the $R^{3+}$ moments in inducing a ferroelectric order in this system. Interestingly, measurements by the same group for o-YMnO$_3$ (Y$^{3+}$ has an empty $f$-shell) revealed a different behavior - $P||a$ showed a significant increase already at the lock-in temperature of the Mn$^{3+}$ moments and its amplitude was larger than that of o-HoMnO$_3$. The appearance of $P_a$ in o-YMnO$_3$, however, is not understood, since the sinusoidal magnetic order, which was reported for this material, should provide zero polarization within the framework of the aforementioned mechanisms for inducing $P$.  Later measurements of $P$ in o-HoMnO$_3$ by Feng \textit{et al.} showed a similar temperature dependence of $P$ to that observed in Ref. \onlinecite{lorenz2007homno3_ymno3} for o-YMnO$_3$, but not for o-HoMnO$_3$. Notably, another measurement of the electric polarization in a sample of o-HoMnO$_3$ with IC magnetic order ($q_b$$\approx$0.4) gave $P$ aligned along the $c$ axis \cite{lee2011homno3}, and for o-ErMnO$_3$ with a similar magnetic ordering no sizable $P$ was observed \cite{ye2007ermno3}.  

Therefore, to better understand the magnetism and ferroelectricity in o-$R$MnO$_3$ and their cross-coupling, it is important to clarify the origin of the inconsistencies described above, to determine possible magnetic phases in these materials and the mechanism of their establishment, and to define how these magnetic phases can induce an electric polarization. In this article we present a detailed analysis of the magnetic and ferroelectric properties of o-HoMnO$_3$ and o-ErMnO$_3$, for which many contradictory results have been reported. 
\subsection{Model Hamiltonian}
\label{sec:hamiltonian}
We study the magnetism in o-$R$MnO$_3$ based on the following model Hamiltonian:
\begin{equation}
\label{fullHam}
H=H_{Heis}+H_{BQ}+H_{4sp}+H_{SIA}+H_{DM},
\end{equation}
where 
\begin{equation}
H_{Heis}=\sum_{<i,j>}J_{ij}\left(\mathbf{S}_i\cdot\mathbf{S}_j\right), 
\label{HeisHam}
\end{equation}
\begin{equation}
H_{BQ}=\sum_{<i,j>}B_{ij}\left(\mathbf{S}_i\cdot\mathbf{S}_j\right)^2,
\label{biqHam}
\end{equation}
\begin{eqnarray}
H_{4sp}=\sum_{<i,j,k,l>}K_{ijkl}\left[ \left(\mathbf {S}_i\cdot\mathbf {S}_j\right)\left(\mathbf S_k\cdot\mathbf S_l\right) \right. \nonumber \\ + \left. \left(\mathbf S_i\cdot\mathbf S_l\right)\left(\mathbf S_k\cdot\mathbf S_j\right) - \left(\mathbf S_i\cdot\mathbf S_k\right)\left(\mathbf S_j\cdot\mathbf S_l\right)\right], 
\label{4bodyHam}
\end{eqnarray}
\begin{equation}
H_{SIA}=A\sum_{i}S^2_{i,b} , 
\label{siaHam}
\end{equation}
\begin{equation}
H_{DM}=\sum_{<i,j>}\mathbf{D}_{ij}\cdot\left[\mathbf{S}_i\times\mathbf{S}_j\right] \quad.
\label{dmHam}
\end{equation}

It includes the following terms: (i) the usual Heisenberg Hamiltonian (Eq.\ \ref{HeisHam}). In its simplest form (including only AFM $J_c$ and $J_b$ and FM $J_{ab}$, see Fig.\ \ref{fig:exchanges}), the Heisenberg Hamiltonian can explain the establishment of the A-AFM order (if $J_b$/$|J_{ab}|$<0.5) and the evolution of the magnetic phase to the IC spiral (for $J_b$/$|J_{ab}|$>0.5) \cite{mochizuki2011theory}. In addition to these three couplings, we include the Heisenberg interactions up to third NN within the $ab$ planes ($J_a$ and $J_{3nn}$) and second NN along the $c$ axis ($J_{diag}$), see Fig.\ \ref{fig:exchanges}. The further-neighbor couplings are not taken into account, since they were found to be negligible in our previous work \cite{fedorova2015biquadratic}; (ii) Biquadratic exchange interactions (Eq.\ \ref{biqHam}). It was demonstrated that the in-plane biquadratic couplings play a crucial role in the establishment of E-AFM order \cite{kaplan2009biquadratic,kaplan2009comment}. In the pure Heisenberg model E-AFM order is degenerate with a 90$^\circ$ spiral, while biquadratic exchange favors magnetic phases with collinear spin orientations. We include in our model the NN biquadratic couplings $B_{ab}$ and $B_c$ (see Fig.\ \ref{fig:exchanges}); (iii) Four-spin ring exchange interactions (Eq.\ \ref{4bodyHam}). We showed in our previous work \cite{fedorova2015biquadratic} that the energies of o-$R$MnO$_3$ calculated using DFT cannot be accurately fitted to the isotropic model Hamiltonian which includes only the Heisenberg and biquadratic couplings, while the addition of the four-spin ring terms significantly improves the fitting. The possible effects of these interactions on the magnetic order in o-$R$MnO$_3$, however, were not investigated before to the best of our knowledge. Here we consider the exchanges between spins in plaquettes within the $ab$ planes ($K_{ab}$) and interplane ones ($K_c$) as shown in Fig.\ \ref{fig:exchanges}; (iv) Single-ion anisotropy (SIA), Eq.\ \ref{siaHam}, which corresponds to the magnetic easy axis along the $b$ direction.
(v) Dzyaloshinskii-Moriya interactions (DMI). We consider DM vectors $\mathbf{D}_{ij}$ defined along Mn-O-Mn bonds in the $ab$ planes ($\mathbf{D}_{ij}^{ab}$) and along the $c$ direction ($\mathbf{D}_{ij}^{c}$). 
It was demonstrated in Ref. \onlinecite{solovyev1996lamno3}, that due to the o-$R$MnO$_3$ crystal symmetry, the DM vectors can be represented in terms of five parameters: $\alpha_{ab}$, $\beta_{ab}$ and $\gamma_{ab}$ for $\mathbf{D}_{ij}^{ab}$ and $\alpha_c$ and $\beta_c$ for $\mathbf{D}_{ij}^{c}$.  It was shown that the $\alpha_c$ component of the DM vectors $\mathbf{D}_{ij}^{c}$ (see Fig.\ 3 in Ref.\ \onlinecite{solovyev1996lamno3}) causes the small canting of the spins from the $b$ axis towards the $c$ axis, which was observed experimentally for several representatives of the o-$R$MnO$_3$ series \cite{matsumoto1970lamno3,mukherjee2017lumno3}. In turn, the $\gamma_{ab}$ components of the $\mathbf{D}_{ij}^{ab}$ vectors give rise to cantings of the Mn$^{3+}$ spins towards the $a$ axis  and also favor the establishment of the $ab$ spiral state \cite{mochizuki2009microscopic}. To enable these states in our simulations, we consider the parameters $\alpha_c$ and $\gamma_{ab}$ to be nonzero, while neglecting the other DM parameters.  

As will be described in detail below, we extract all the exchange coupling and anisotropy constants for o-HoMnO$_3$ and o-ErMnO$_3$ by mapping the results of \textit{ab initio} electronic structure calculations onto this model Hamiltonian (Eq.\ \ref{fullHam}) and perform a series of MC simulations using the obtained couplings to determine the corresponding ground states of this Hamiltonian.

\section{Computational details}
\label{sec:comput_details}
All \textit{ab initio} electronic structure calculations are performed using the projector-augmented plane-wave method of DFT \cite{hohenberg1964DFT,kohn1965DFT} as implemented in the Vienna \textit{ab initio} Simulation Package (VASP) \cite{kresseVasp}. We employ the generalized gradient approximation plus Hubbard $U$ (GGA+$U$) for the exchange-correlation potential in the form introduced by Perdew, Burke and Ernzerhof in the version revised for solids (PBEsol) \cite{perdew2008Pbesol}. In this study we do not consider effects which may originate from ordering of the $f$-electron moments of the $R$ cations, therefore we use  pseupodentials for $R$ elements in which $f$ electrons are treated as core electrons. The parameter of on-site Coulomb repulsion $U$ for the $d$ states of Mn is set to 1 eV as it gives a reasonable size of the band gaps and correct magnetic ground states for many o-$R$MnO$_3$. The cutoff energy for the plane wave basis set is 600 eV. All the calculations with 20-atom unit cells are done using a $\Gamma$-centered 7$\times$7$\times$5 k-point mesh. For 80-atom  2$\times$2$\times$1 supercells (obtained by doubling the 20-atom unit cell along the crystallographic $a$ and $b$ directions) we choose a 3$\times$3$\times$5 k-point mesh, while for 1$\times$2$\times$2 supercells (with the 20-atom unit cell doubled along the $b$ and $c$ axes) we use a 7$\times$3$\times$2 k-point mesh.

Unless otherwise specified, the structural optimizations are performed using 20-atom unit cells and imposing A-AFM order of  the Mn spins. The structure is considered to be relaxed when the Hellmann-Feynman forces acting on the atoms are below 10$^{-4}$ eV/\AA. In cases when the volume is allowed to relax, we ensure that the components of the stress tensor are smaller than 0.1 kbar.

Monte Carlo simulations are performed using an internally developed code based on the Metropolis algorithm \cite{metropolis1953montecarlo} combined with over-relaxation moves \cite{creutz1987overrelaxation}. Since we are dealing with systems with many competing exchange interactions, which give a complex free energy landscape, we employ the replica exchange technique \cite{swendsen1986replica_exchange,earl2005partemp}. For every system we simulate in parallel $M$=200 replicas, each at a different temperature. Temperatures are distributed exponentially with $T_k$=$T_0/\alpha^k$, where $T_0$=0.005 meV is a temperature of interest, $k$=1...$M-1$ and $\alpha$=0.962 (this value is chosen so that the highest temperature $T_{M-1}$ is bigger than the absolute value of the strongest exchange interactions in the considered systems). 
Unless otherwise specified, we perform simulations with 12$\times$40$\times$12 unit cells and 4$\times$100$\times$4 unit cells each containing 2 Mn atoms: Mn$_1$ (0,0.5,0) and Mn$_2$ (0.5,1,0) and apply periodic boundary conditions in all directions. We repeat the calculations using open boundary conditions along the $b$ direction to ensure that the choice of boundary conditions does not affect the (in)commensurability of the obtained magnetic ground states. Since we work with systems with numerous competing exchange interactions, we perform calculations starting from different types of magnetic order (A-AFM, E-AFM, random orientation and H-AFM, the latter will be described below) to ensure that the results are not affected by the starting configurations and the systems are not trapped in a local energy minimum.

\section{Results and discussion}
\label{sec:results}
\subsection{HoMnO$_3$} 
\subsubsection{Magnetic order}
\label{subsec:results_ho_magn}
First we investigate the magnetic properties of o-HoMnO$_3$.
We begin our analysis by optimizing the volume and ionic positions of the bulk crystal structure of o-HoMnO$_3$ using DFT.
We start from the experimentally reported structure \cite{lee2011homno3} for which incommensurate magnetic order with $q_b$$\approx$0.4 was observed. The obtained lattice parameters together with the experimental data are summarized in Table \ref{tab:lattice_parameters}. 
\begin{table}[b]
\caption{The experimental and theoretically optimized lattice parameters of bulk o-HoMnO$_3$ and o-ErMnO$_3$. $a$, $b$ and $c$ are the lattice constants (space group \textit{Pbnm}, \#62); $s$, $m$ and $l$ are the short, medium and long Mn-O bonds in the MnO$_6$ octahedra; IPA and OPA are the Mn-O-Mn bond angles within the $ab$ planes and along the $c$ direction, respectively. All distances are in \AA, all angles are in degrees.}
\begin{tabular}{p{41pt}p{46pt}p{46pt}|p{46pt}p{46pt}
}
\hline
\hline
& \multicolumn{2}{c}{HoMnO$_3$} & \multicolumn{2}{c}{ErMnO$_3$}
\tabularnewline
& \multicolumn{2}{c}{powder} & \multicolumn{2}{c}{powder} 
\tabularnewline
& \centering{Exp\cite{lee2011homno3}} & \centering{PBE} & \centering{Exp\cite{ye2007ermno3}} & \centering{PBE}  
\tabularnewline
\hline
\centering{$a$} & \centering{5.269} & \centering{5.203} & \centering{5.227} & \centering{5.186} 
\tabularnewline
\centering{$b$} & \centering{5.845} & \centering{5.772} & \centering{5.792} & \centering{5.759} 
\tabularnewline
\centering{$c$} & \centering{7.370} & \centering{7.301} & \centering{7.327} & \centering{7.282} 
\tabularnewline
\centering{$s$} & \centering{-} & \centering{1.913} & \centering{1.910} & \centering{1.911} 
\tabularnewline
\centering{$m$} & \centering{-} & \centering{1.936} & \centering{1.938} & \centering{1.936} 
\tabularnewline
\centering{$l$} & \centering{-} & \centering{2.174} & \centering{2.194} & \centering{2.170} 
\tabularnewline
\centering{IPA} & \centering{-} & \centering{143.85} & \centering{143.66} & \centering{143.32} 
\tabularnewline
\centering{OPA}& \centering{-} & \centering{141.08} & \centering{141.91} & \centering{140.15} 
\tabularnewline

\hline
\hline
\end{tabular}
\label{tab:lattice_parameters}
\end{table} 
\begin{table*}[htb]
\caption{Calculated Heisenberg ($J_c$, $J_{ab}$, $J_a$, $J_{diag}$, $J_b$, $J_{3nn}$), four-spin ring ($K_{ab}$ and $K_c$) and biquadratic ($B_{ab}$ and $B_c$) exchanges, components of DM vectors ($\gamma_{ab}$ and $\alpha_c$) and single-ion anisotropies ($A$) (in meV) in bulk o-HoMnO$_3$ and o-ErMnO$_3$.}
\begin{tabular}{p{31pt}p{32.5pt}p{32.5pt}p{32.5pt}p{32.5pt}p{32.5pt}p{32.5pt}p{32.5pt}p{32.5pt}p{32.5pt}p{32.5pt}p{32.5pt}p{32.5pt}p{32.5pt}p{32.5pt}}
\hline
\hline
\centering{$R$} & \centering{$J_c$} & \centering{$J_{ab}$} & \centering{$J_a$} &\centering{$J_{diag}$} & \centering{$J_b$} & \centering{$J_{3nn}$} & \centering{$K_{ab}$} & \centering{$K_c$} & \centering{$B_{ab}$} & \centering{$B_c$} & \centering{$\gamma_{ab}$} & \centering{$\alpha_c$} & \centering{$A$}       
\tabularnewline
\hline 
\centering{Ho} & \centering{4.23} & \centering{-4.49} & \centering{-1.02} & \centering{0.70} & \centering{0.87} & \centering{2.69} &\centering{0.29} & \centering{0.90} & \centering{-2.27} & \centering{-0.51} & \centering{-0.57} & \centering{-0.42} & \centering{-0.48}
\tabularnewline
\centering{Er} & \centering{4.20} & \centering{-3.81} & \centering{-0.99} & \centering{0.69} & \centering{0.95} & \centering{2.68} &\centering{0.28} & \centering{0.90} & \centering{-2.25} & \centering{-0.45} & \centering{-0.58} & \centering{-0.40} & \centering{-0.49}
\tabularnewline
\hline
\hline
\end{tabular}
\label{tab:allcouplings}
\end{table*}
\begin{figure*}
\includegraphics[width=0.95\textwidth]{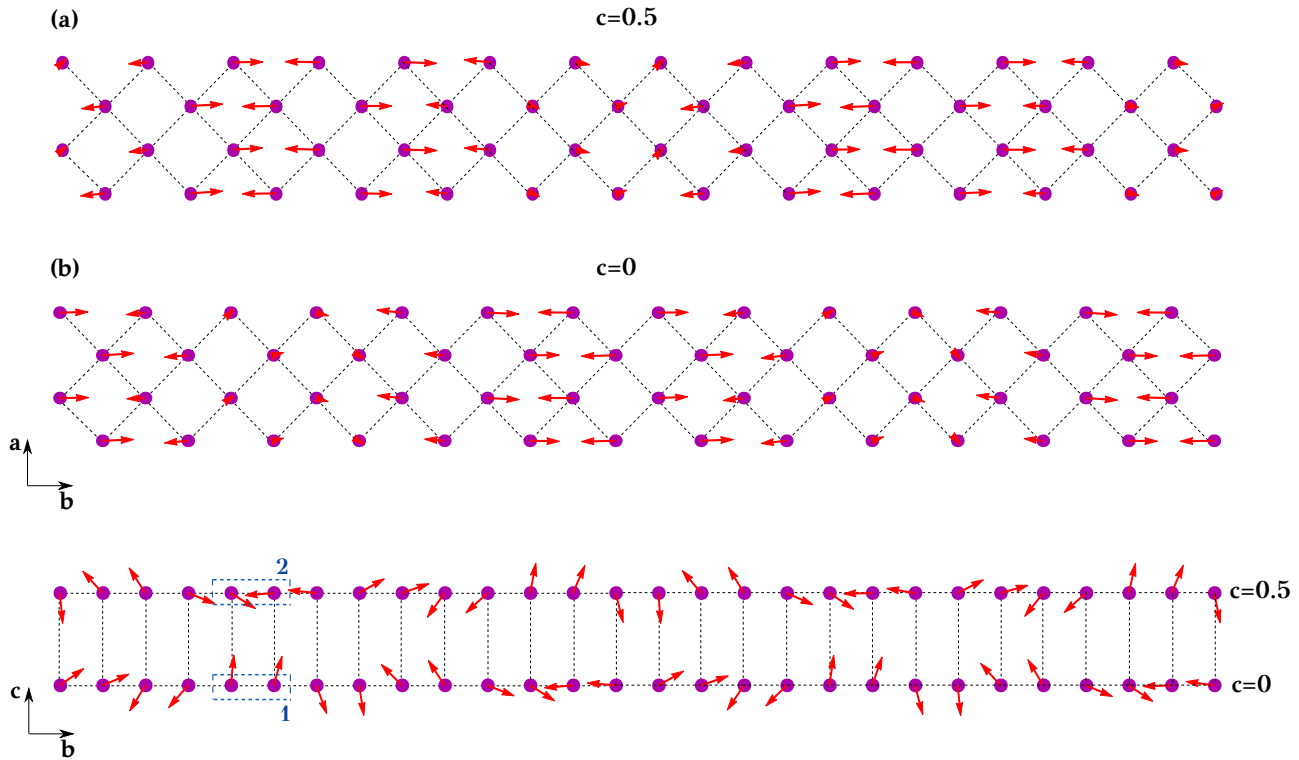}
\caption{\label{fig:wSpiral} Magnetic ground state (w-spiral) obtained using MC simulations for o-HoMnO$_3$ (system with 2$\times$14$\times$1 crystallographic unit cells is shown, purple spheres denote the Mn ions). (a) and (b) show the spins (red arrows) in the $ab$ planes at $c$=0.5 and $c$=0, respectively; (c) illustrates the orientation of spins within the $bc$ planes.}
\end{figure*}
We calculate all the couplings described above by mapping the results of DFT calculations onto the considered model Hamiltonian (see Eq.\ \ref{fullHam}). First we extract the Heisenberg and four-spin ring couplings by constructing an 80 atom supercell and calculating the energies of $N_{eq}$=32 inequivalent collinear magnetic orders (all the states are insulating). Note that the spin-orbit coupling is not included in these calculations (DMI and SIA are excluded). 
We use the obtained energies to construct an overdetermined system of equations, in which the left-hand side of each equation is written using Eqs. \ref{HeisHam} and \ref{4bodyHam} and the coupling constants $J_c$, $J_{ab}$, $J_a$, $J_{diag}$, $J_b$, $J_{3nn}$, $K_{ab}$ and $K_c$ (see Fig.\ \ref{fig:exchanges}) are unknowns. The lowest energy configuration (E-AFM order) is taken as a reference. We solve the system of equations using the least mean square method and find all the couplings mentioned above (see Table \ref{tab:allcouplings}). We note that the extracted couplings can be affected by the set of considered equations. Generally speaking, when one has a relatively large number of equations $N_{eq}$ in the system, adding or removing one or more does not change the resulting coupling constants much. Nevertheless, as one can see from Appendix \ref{app:exchanges}, an uncertainty of up to $\approx\pm25\%$ from the values presented in Table \ref{tab:allcouplings} is possible.

In the next step we extract the biquadratic couplings $B_{ab}$ and $B_c$ using the approach described in detail in Sec.\ IVB of our previous work  \cite{fedorova2015biquadratic}. We perform a set of calculations of the total energies $E$ of o-HoMnO$_3$ using a 20-atom unit cell in which one of the Mn spins is rotated by an angle $\alpha$ from 0 to 180$^\circ$ starting from a specific noncollinear spin state. The couplings $B_c$ and $B_{ab}$ are obtained from fitting the calculated $E(\alpha)$ to the functions $f(\alpha)$=$C_1+C_2\cos(\alpha)+C_3\cos^2(\alpha)$. Note  that this method takes the $C_2$ parameters extracted from the fitting to the Heisenberg Hamiltonian so that the uncertainty we mentioned previously in the $C_2$ values propagates into the biquadratic couplings, which are a part of the parameter $C_3$ (it also contains the four-spin ring part, which is known from the previous calculations). 
To extract the components of the DM vectors ($\gamma_{ab}$ and $\alpha_c$) and the SIA ($A$), we employ the method proposed in Sec.\ IIC of Ref.\ \onlinecite{xiang2011whangbo}. We perform calculations of the energies of a 20-atom unit cell with noncollinear magnetic orders including spin-orbit coupling. All obtained coupling constants are summarized in Table \ref{tab:allcouplings}. 
One can see that almost all the couplings which we extracted using DFT for o-HoMnO$_3$ are relatively strong and competition between them may result in magnetic frustration.  

In order to determine the ground state of our Hamiltonian (Eq.\ \ref{fullHam}) for o-HoMnO$_3$, we perform a series of MC simulations using the calculated exchange couplings listed in Table \ref{tab:allcouplings}. 
We define the type of the resulting magnetic orders based on the calculations of the following quantities: order parameters for A-AFM, E-AFM, H-AFM and I-AFM states (see Eqs.\ \ref{eq:OPA}-\ref{eq:OPsW4} in Appendix, H-AFM and I-AFM orders will be described later in this section) and magnetic structure factors (Eqs.\ \ref{eq:struct_fact1} and \ref{eq:struct_fact2}) along different directions in reciprocal space.
We find that the ground-state ordering of Mn spins in o-HoMnO$_3$ is the E-AFM order with a propagation vector $\mathbf{q}$=(0,0.5,0). The spins are mostly aligned along the $b$ axis and have negligible $a$ and $c$ components which are favored by the DMI. It should be noted, however, that the only simulations which converge to this state are those starting from perfect E-AFM order, while simulations using other magnetic orders as starting configurations give states corresponding to local energy minima with slightly higher energies. Such a behavior was not observed in MC simulations that we performed for other representatives of the o-$R$MnO$_3$ series such as GdMnO$_3$ and TbMnO$_3$ (both bulk and thin films). This is an indication of competition between different exchange interactions in o-HoMnO$_3$ resulting in multiple magnetic states with very close energies, which makes it very hard to find a global energy minimum in MC simulations. 

\begin{figure}
\begin{centering}
\includegraphics[width=0.41\textwidth]{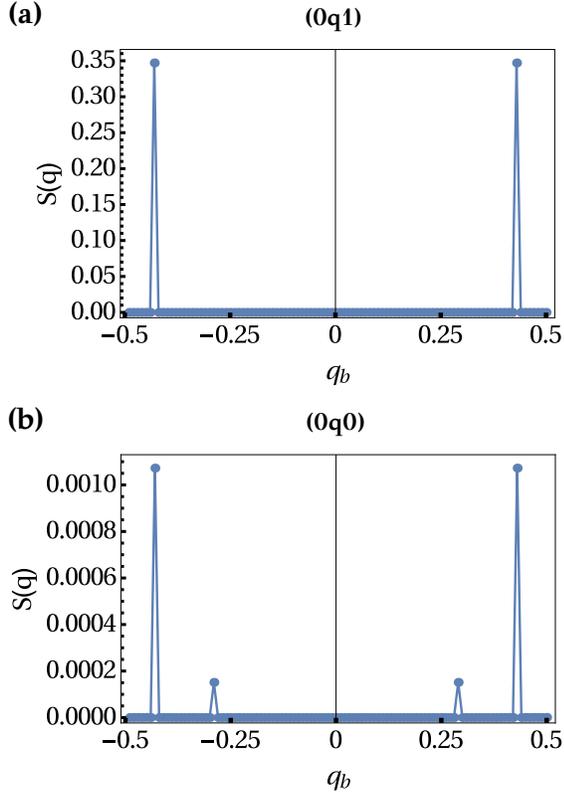}
\caption{\label{fig:wSpiralSF} Magnetic structure factors along (a) (0,$q$,1) and (b) (0,$q$,0) directions in reciprocal space calculated for the w-spiral ordering obtained in MC simulations for o-HoMnO$_3$. $q_b$ is in reciprocal lattice units.}
\end{centering}
\end{figure}

As we mentioned above, the methods which we use to extract the bilinear, biquadratic and four-spin ring exchanges allow an uncertainty of up to $\pm25\%$ for each considered coupling. In order to check whether such a variation of the exchange couplings may lead to different magnetic ground states, we perform the following analysis. We repeat the MC simulations using the set of couplings, obtained by solving the overdetermined system of equations with respect to all bilinear and four-spin ring couplings, that differs most strongly from those obtained with $N_{eq}$=32 (see the couplings corresponding to $N_{eq}$=33 in Table \ref{tab:couplings_vs_equations} of Appendix \ref{app:exchanges}). We find that this set of couplings gives a different magnetic ground state for o-HoMnO$_3$. This state, which we call w-spiral order, is shown in Fig.\ \ref{fig:wSpiral}.
The magnetic structure factors calculated for this order along (0,$q$,1) and (0,$q$,0) directions in reciprocal space are shown in Fig.\ \ref{fig:wSpiralSF} (a) and (b), respectively. One can see that this order gives peaks at (0,$\pm$0.43,1) corresponding to a propagation vector of $\mathbf{q}$=(0,0.43,0), which is similar to that reported from experiment ($q_b\sim0.4$) \cite{brinks2001homno3,lee2011homno3}. We find also that it produces peaks at (0,$\pm$0.43,0) and (0,$\pm$0.29,0), with much smaller intensities than that at (0,$\pm q_b$,1).

\begin{table}[!b]
\caption{\label{tab:variationCouplings} Magnetic ground states obtained using MC simulations for o-HoMnO$_3$ and o-ErMnO$_3$ by varying one of the exchange couplings by $\pm30\%$ with all other couplings kept fixed to those listed in Table \ref{tab:allcouplings}. The variation of $J_a$, $J_{diag}$, $K_{ab}$, $B_{ab}$, $B_c$, $\gamma_{ab}$, $\alpha_c$ and $A$ did not lead to a change of the ground state (E-AFM remains), therefore they are omitted in this table.} 

\begin{tabular}
{p{35pt}p{100pt}p{100pt}}
\hline
\hline
\centering{$J$} & \centering{HoMnO$_3$} & \centering{ErMnO$_3$}
\tabularnewline
\hline
\multirow{2}{35pt}{\centering{$J_c$}} & \centering{-30\% \\ H-AFM $q_b$=0.5} & \centering{-30\% \\ H-AFM $q_b$=0.5}
\tabularnewline
\multirow{2}{35pt}{\centering{$J_{ab}$}} & \centering{+30\% \\ Spiral $q_b$=0.18} & \multirow{2}{100pt}{\centering{-}}
\tabularnewline
\multirow{2}{35pt}{\centering{$J_{3nn}$}} & \multirow{2}{100pt}{\centering{-}} & \centering{-30\% \\ A-AFM}
\tabularnewline
\multirow{2}{35pt}{\centering{$K_c$}} & \centering{+30\% \\ w-spiral $q_b$=0.43} & \centering{+30\% \\ w-spiral $q_b$=0.45}
\tabularnewline
\hline
\hline
\end{tabular}
\end{table}

\begin{figure}
\begin{centering}
\includegraphics[width=0.32\textwidth]{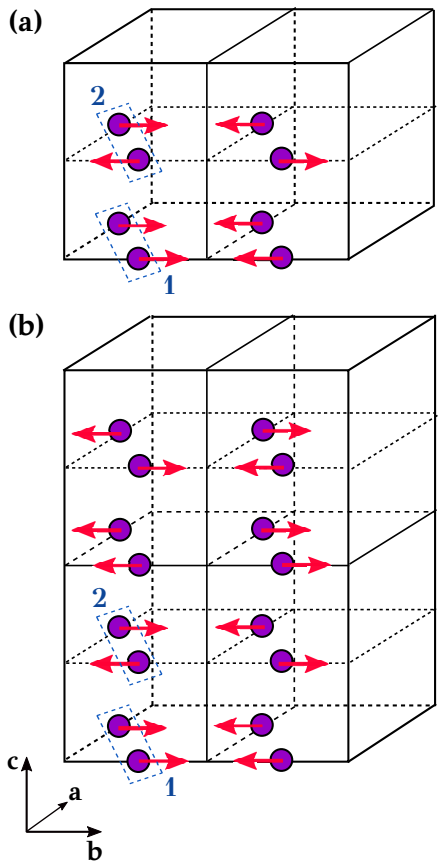}
\caption{\label{fig:weird_sW} Magnetic orders that can be stabilized in o-HoMnO$_3$ by strong interplane four-spin ring exchange coupling $K_c$. (a) shows a magnetic unit cell ($1\times2\times1$ crystallographic unit cells) of so-called H-AFM order with $\mathbf{q}$=(0,0.5,0); (b) is a magnetic unit cell ($1\times2\times2$ crystallographic unit cells) of so-called I-AFM order with a propagation vector $\mathbf{q}=(0,0.5,0.5)$.  Purple spheres denote Mn ions and red arrows show the spins of these ions. 
}
\end{centering}
\end{figure}

\begin{figure}
\begin{centering}
\includegraphics[width=0.41\textwidth]{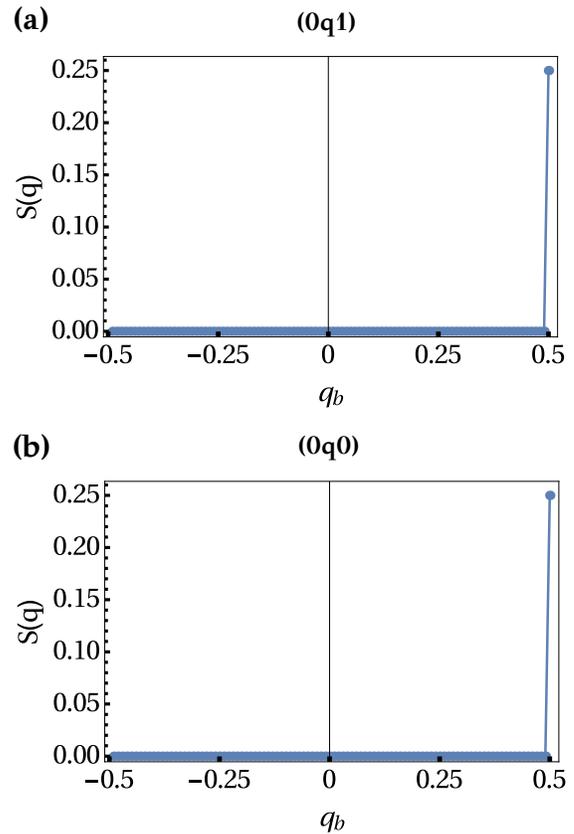}
\caption{\label{fig:Sw_SF_HoMnO3} Magnetic structure factors (a) along (0,$q$,1) direction and (b) along (0,$q$,0) direction calculated for the H-AFM ordering observed in MC simulations for o-HoMnO$_3$ using $J_c$=$0.7J_c^0$, where $J_c^0$ is the value presented in Table \ref{tab:allcouplings}, while all the other couplings are kept fixed to those in Table \ref{tab:allcouplings}. $q_b$ is in reciprocal lattice units.}
\end{centering}
\end{figure}

Next, in order to identify which terms in the Hamiltonian of Eq.\ \ref{fullHam} are responsible for the establishment of the exotic w-spiral state, we perform a series of MC simulations in which one of the exchange couplings is varied by $\pm30\%$ while all the others are kept fixed to those listed in Table \ref{tab:allcouplings}. We run the simulations with a system size of 4$\times$100$\times$4 MC unit cells (Fig.\ \ref{fig:MC_unitcell}) and employ periodic boundary conditions in all directions. Since the exchange couplings in Table \ref{tab:allcouplings} give E-AFM order as the ground state, we use this order as the starting configuration in all the simulations with varied couplings. The results of our simulations are summarized in Table \ref{tab:variationCouplings}. From this one can see that variation of three exchange couplings may lead to stabilization of a magnetic state in o-HoMnO$_3$ different from the E-AFM order: (i) increase of the four-spin ring exchange $K_c$ results in the establishment of the w-spiral order with $q_b$=0.43 shown in Fig. \ref{fig:wSpiral};
(ii) increasing the NN in-plane Heisenberg coupling $J_{ab}$ gives a spiral ordering with a propagation vector along the $b$ axis $q_b$=0.18 (a state with such a propagation vector, however, has not been reported for o-HoMnO$_3$ to the best of our knowledge); (iii) 
reduction of the NN Heisenberg coupling $J_c$ favors an order which we call H-AFM. A sketch of this order is shown in Fig. \ref{fig:weird_sW} (a).  It has a propagation vector $\mathbf{q}=(0,0.5,0)$ and gives peaks in the magnetic structure factors at (0,$\pm$0.5,1) and (0,$\pm$0.5,0) with the same intensity (see Fig.\ \ref{fig:Sw_SF_HoMnO3}).  
\begin{table}[b]
\caption{DFT energies per spin (in meV) (relative to the energy of the E-AFM order) and electric polarizations (in $\mu$C/cm$^2$) calculated for o-HoMnO$_3$ and o-ErMnO$_3$ imposing E-AFM, H-AFM and I-AFM orders.}
\begin{tabular}{p{36pt}p{48pt}p{48pt}|p{48pt}p{48pt}}
\hline
\hline
& \multicolumn{2}{c}{HoMnO$_3$} & \multicolumn{2}{c}{ErMnO$_3$} 
\tabularnewline
& \centering{$E$} & \centering{$P$} & \centering{$E$} & \centering{$P$} 
\tabularnewline
\hline
\centering{E-AFM} & \centering{0} & \centering{4.09, || $a$} & \centering{0} & \centering{4.06, || $a$} 
\tabularnewline
\centering{H-AFM} & \centering{2.19} & \centering{0.34, || $c$} & \centering{2.13} & \centering{0.35, || $c$} 
\tabularnewline
\centering{I-AFM} & \centering{1.38} & \centering{0.12, || $a$} & \centering{1.38} & \centering{0.12, || $a$}  
\tabularnewline
\hline
\hline
\end{tabular}
\label{table_DFT_en}
\end{table} 
\begin{table}[t]
\caption{\label{tab:compStatesHo} Exact energies per spin (in meV) of A-AFM, E-AFM, $bc$ and $ab$ cycloidal spirals with $q_b$=0.43 (CS$_{bc}$ and CS$_{ab}$), $bc$ cycloidal spiral with $q_b$=0.18, w-spiral, H-AFM (energy is equal to I-AFM) and sinusoidal (SIN) orders calculated for o-HoMnO$_3$ and o-ErMnO$_3$ using Eq.\ \ref{fullHam} and  exchange couplings extracted from DFT and listed in Table \ref{tab:allcouplings}.}
\begin{tabular}
{p{72pt}p{81pt}p{81pt}}
\hline
\hline
\centering{State} & \centering{HoMnO$_3$} & \centering{ErMnO$_3$}
\tabularnewline
\hline
\centering{A-AFM} & \centering{-14.197} & \centering{-12.629}
\tabularnewline
\centering{E-AFM} & \centering{-15.523} & \centering{-15.418}
\tabularnewline
\centering{CS$_{bc}$ $q_b$=0.43} & \centering{-12.224} & \centering{-11.870}
\tabularnewline
\centering{CS$_{ab}$ $q_b$=0.43} & \centering{-12.223} & \centering{-11.869}
\tabularnewline
\centering{CS$_{bc}$ $q_b$=0.18} & \centering{-14.439} & \centering{-}
\tabularnewline
\centering{w-spiral} & \centering{-15.177} & \centering{-14.999}
\tabularnewline
\centering{H-AFM} & \centering{-14.884} & \centering{-14.832}
\tabularnewline
\centering{SIN} & \centering{-7.533} & \centering{-7.364}
\tabularnewline
\hline
\hline
\end{tabular}
\end{table}
Notably, the H-AFM magnetic state is degenerate (if we calculate its energy using Eq.\ \ref{fullHam}) with another order, which we call I-AFM. This order has a propagation vector $\mathbf{q}=(0,0.5,0.5)$ and is shown in Fig.\ \ref{fig:weird_sW} (b). A similar order with $\mathbf{q}$=(0.5,0,0.5) was reported from neutron diffraction experiments on the o-$R$NiO$_3$ series\cite{alonso2001rnio3,garcia1994rnio3}. This order does not give peaks in the magnetic structure factors at either (0,$q$,1) or (0,$q$,0).  To check whether the I-AFM state might be favored over H-AFM (or vice versa) in o-HoMnO$_3$ due to effects such as exchange striction or distortion of the electronic density, we perform the following DFT calculations: We construct a 1$\times$2$\times$2 supercell of o-HoMnO$_3$ (the fully relaxed unit cell of o-HoMnO$_3$ is doubled along the $b$ and $c$ directions) and relax the ionic positions within this supercell imposing  E-AFM, H-AFM and I-AFM orders. After that we use these optimized structures to calculate the total energies of o-HoMnO$_3$ with corresponding magnetic orders. The obtained energies with respect to the energy of the E-AFM order (the lowest energy state for o-HoMnO$_3$ in DFT) are presented in Table \ref{table_DFT_en}. We find that I-AFM order is lower in energy than H-AFM by $\approx$0.8 meV per spin. Note, that all three orders (E-, H- and I-AFM) are very close in energy and we assume that any of them may, in principle, be stabilized in real samples of o-$R$MnO$_3$. Favoring one of these states over the others can occur due to different synthesis conditions which may provide slightly different bond angles and bond lengths and, therefore, different exchange couplings in the systems.

The fact that the variation of the interplane couplings may drastically change the magnetic ground state is interesting in its own right, as in all previous works the role of the interplane couplings was considered only to explain an antiferromagnetic orientation of spins along the $c$ direction. Furthermore, we see that the w-spiral, H-AFM and I-AFM (Fig.\ \ref{fig:wSpiral}, \ref{fig:weird_sW} (a) and \ref{fig:weird_sW} (b), respectively) orders minimize the energy contribution from the four-spin ring interplane exchange (Eq.\ \ref{4bodyHam}). Indeed, for H-AFM and I-AFM orders each interplane four-site plaquette contains one pair of spins which are parallel to each other (pair 1) and one pair in which the spins are antiparallel to each other (pair 2), which gives a contribution to the energy of -2$K_c$ per spin according to Eq.\ \ref{4bodyHam}. For the w-spiral order the spins of pair 1 (see Fig. \ref{fig:wSpiral}) are almost parallel to each other, while the spins of pair 2 are almost antiparallel to each other and the spins of pair 1 and pair 2 are nearly perpendicularly oriented. This also gives a contribution between -$K_c$ and -2$K_c$ per spin to the total energy. For comparison, for the E-AFM order the contribution to the energy of the system from the four-spin ring exchange is $+2K_c$ per spin (note that the only couplings which give different contributions to the total energies of E-AFM and H-AFM (I-AFM) states according to Eq.\ \ref{fullHam} are $J_c$ and $K_c$ and the contributions due to other exchange interactions or anisotropies are equal). Therefore, w-spiral, H-AFM and I-AFM orders may become the lowest energy states when either $K_c$ is increased or $J_c$ is reduced (AFM $J_c$ cannot further compete with strong $K_c$). Thus we conclude that these orders are favored by the four-spin ring interplane coupling $K_c$. 
\begin{figure}
\begin{centering}
\includegraphics[width=0.41\textwidth]{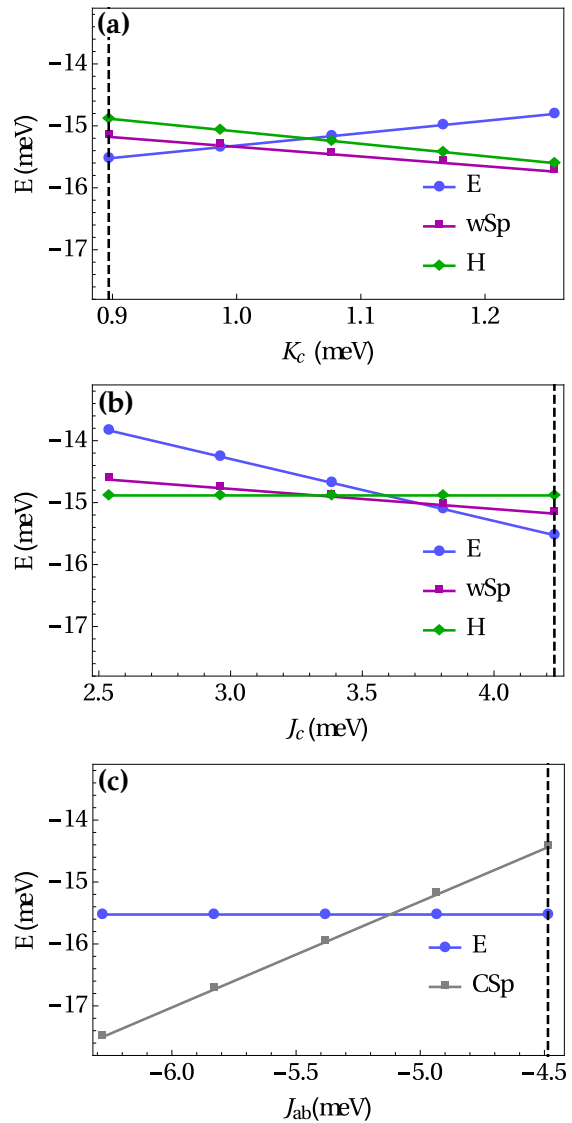}
\caption{\label{fig:en_comp_Ho} Exact energies (per spin) of different magnetic phases of o-HoMnO$_3$ as functions of $K_c$, $J_c$ and $J_{ab}$. (a) shows the energies of the E-AFM (E), w-spiral (wSp) and H-AFM (H) or I-AFM orders as functions of $K_c$; (b) shows the energies of the same orders as functions of $J_c$; (c) demonstrates the energies of the E-AFM and cycloidal spiral (CSp) orders with $q_b$=0.18 as functions of $J_{ab}$. Dashed lines indicate the initial values of the corresponding couplings extracted using DFT and presented in Table \ref{tab:allcouplings}.}
\end{centering}
\end{figure}

Finally, we double check the results of Monte Carlo simulations 
by calculating the exact energies of different magnetic states which may, in principle, occur in o-$R$MnO$_3$ (A-AFM, cycloidal spiral with $q_b$=0.43 and the spins rotating within $ab$ (CS$_{ab}$) and $bc$ (CS$_{bc}$) planes and sinusoidal order (SIN) with $q_b$=0.43 which was described in Ref.\ \onlinecite{brinks2001homno3}) and of those which we found in our MC simulations (E-AFM, H-AFM (or I-AFM),  cycloidal spiral with $q_b$=0.18 and w-spiral with $q_b$=0.43) using Eq.\ \ref{fullHam} and the set of couplings for o-HoMnO$_3$ listed in Table \ref{tab:allcouplings}. The energies per spin obtained for each aforementioned magnetic order are presented in Table \ref{tab:compStatesHo}. One can see that, with the couplings presented in Table \ref{tab:allcouplings}, E-AFM is indeed the lowest energy state compared to all other states listed in Table \ref{tab:compStatesHo}. The w-spiral state is the second lowest in energy and differs from E-AFM by $\approx$0.35 meV per spin. Note that cycloidal spiral orders and sinusoidal order SIN with $q_b$=0.43 are significantly higher in energy than all the other considered orders, therefore we conclude that the w-spiral with $q_b$=0.43 is more likely to form in this system than the other incommensurate states. 

In the next step, we define more precisely the ranges of the three exchange couplings ($K_c$, $J_c$ and $J_{ab}$) in which the transition from E-AFM to another ground state takes place. We calculate the exact energies of E-AFM, H-AFM (I-AFM) and w-spiral orders for the values of four-spin ring coupling $K_c$=$K_c^0$+$i K_c^0$, where $i$=0.1,...,0.4 and $K_c^0$ is the value presented in Table \ref{tab:allcouplings}, keeping all the other couplings fixed to those in Table \ref{tab:allcouplings}. The obtained energies as functions of $K_c$ are presented in Fig.\ \ref{fig:en_comp_Ho} (a). One can see that an increase of $K_c$ by $\approx$12\% (which is just 0.11 meV) favors the establishment of w-spiral order in o-HoMnO$_3$. H-AFM (or I-AFM) order, in turn, is higher in energy than the w-spiral in the whole range of the considered values of $K_c$. 
Then we perform the same calculations varying $J_c$ in the range $\left(J_c^0,J_c^0-0.4J_c^0\right)$ with a step of 0.1$J_c^0$ (see $E(J_c)$ in Fig.\ \ref{fig:en_comp_Ho}(b)). In this case the decrease of the coupling (by $\approx$12\%, which is $\approx$0.51 meV) stabilizes first the w-spiral state and then, when $J_c$ is reduced by more than 20\% ($\approx$0.85 meV),the H-AFM (or I-AFM) order becomes the lowest energy state.  Finally, we compare the energies of the E-AFM order and $bc$ spiral with propagation vector of $q_b$=0.18 for the values of $J_{ab}^0$ from $J_{ab}^0$ to $J_{ab}^0$+0.4$J_{ab}^0$ with a step of 0.1$J_{ab}^0$ and the resulting $E(J_{ab})$ is shown in Fig.\ \ref{fig:en_comp_Ho}(c). 
One can see that the cycloidal spiral with $q_b$=0.18 can become the lowest energy state if the absolute value of $J_{ab}$ is increased by $\approx$15\% (0.7 meV).

\begin{figure*}
\begin{centering}
\includegraphics[width=0.99\textwidth]{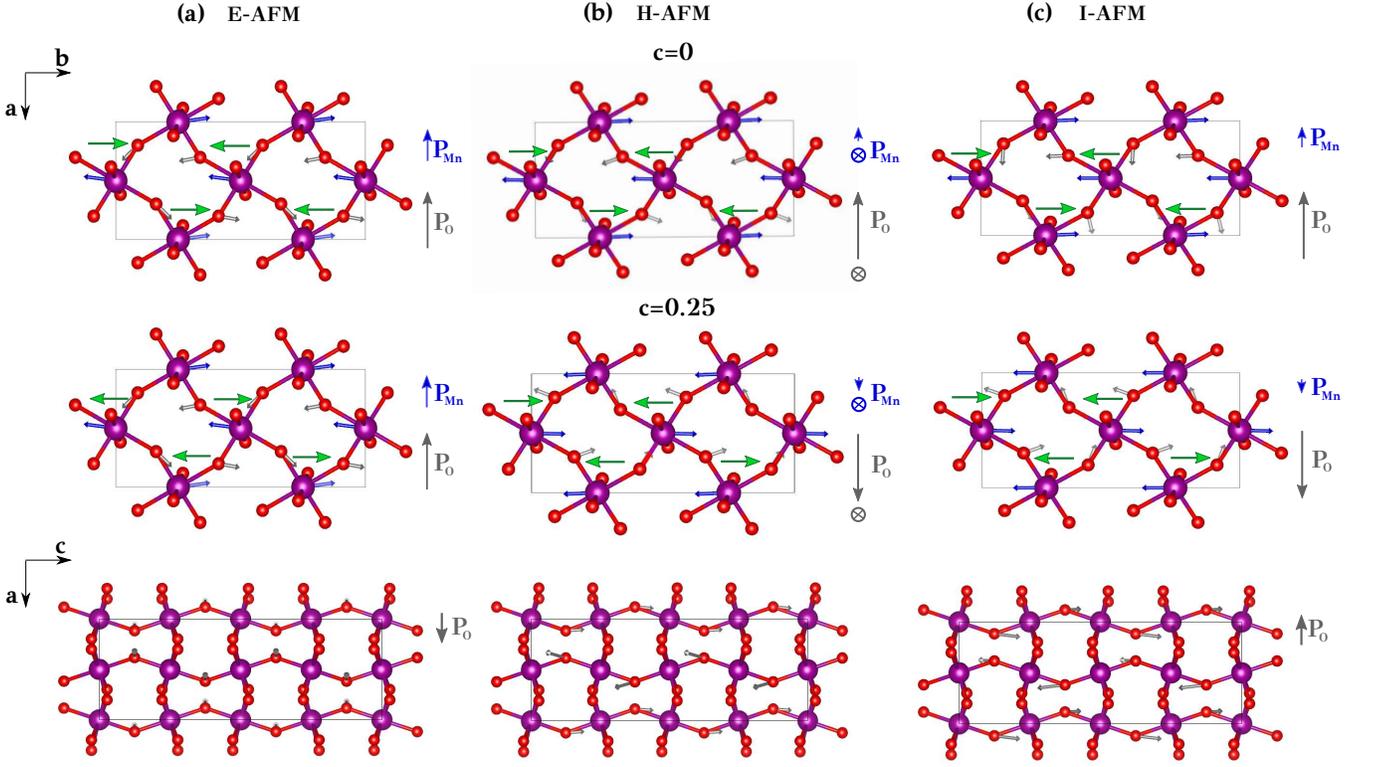}
\caption{\label{fig:polarizations} Electric polarizations due to the displacements of Mn and O ions caused by (a) E-AFM, (b) H-AFM and (c) I-AFM orders. 1$\times$2$\times$2 o-HoMnO$_3$ supercell is shown (only Mn (purple spheres) and O (red spheres) ions). Blue arrows on the Mn ions indicate their displacements, gray arrows - the shifts of the O anions. Green arrows indicate the direction of spins of the Mn ions. $P_{Mn}$ is the contribution to the electric polarization arising from the displacements of the Mn ions, $P_O$ the corresponding contribution from the oxygens. Note, that for the E-AFM and H-AFM orders the directions of the spins at c=0 are equivalent to those at c=0.5 and the same holds for the spin directions at c=0.25 and c=0.75. For I-AFM order, however, the spins at c=0 are antiparallel to the spins at c=0.5 and the spins at c=0.25 are antiparallel to those at c=0.75.}
\end{centering}
\end{figure*}

Thus we find that within the uncertainty of the method which we use to extract the microscopic bilinear and higher order exchange interactions, several magnetic ground states are possible in o-HoMnO$_3$: E-AFM, cycloidal spiral with $q_b$=0.18, w-spiral with $q_b$=0.43 and H-AFM (or I-AFM since it has the same energy as H-AFM within the framework of the considered model, Eq.\ \ref{fullHam}). w-spiral, H-AFM and I-AFM phases are favored by strong interplane four-spin ring exchange $K_c$. Since all these states are very close in energy, one or another could be favored in real samples of o-HoMnO$_3$ depending on the synthesis conditions and the quality of the samples, since slightly different Mn-O bond lengths and/or Mn-O-Mn bond angles in different o-HoMnO$_3$ samples can give different sets of exchange couplings which may favor various magnetic phases. This may explain the experimental observations of both E-AFM order with $q_b$=0.5 \cite{munoz2001homno3} and IC order with $q_b\approx0.4$ \cite{brinks2001homno3,lee2011homno3} in different samples of o-HoMnO$_3$. 
We find, however, that the IC phase is more likely to be a w-spiral state than a sinusoidal spin density wave, since the former is significantly lower in energy than the latter within the framework of the considered model (Eq.\ \ref{fullHam}). To verify this, further experimental investigations or reconsideration of the existing data are required. 

\subsubsection{Ferroelectric order}
\begin{table*}[htb]
\caption{Displacements of the Mn and O ions within the 1$\times$2$\times$2 supercells of o-HoMnO$_3$ and o-ErMnO$_3$ from their positions in the corresponding centrosymmetric structures due to the presence of the E-AFM, H-AFM and I-AFM orders. O$^{ab}_p$ indicates oxygen ions placed between the NN Mn ions with parallel spins within the $ab$ planes, O$^{ab}_{ap}$ denotes the oxygen ions between the NN Mn ions with antiparallel spins within the $ab$ planes. O$^{c}_p$ and O$^{c}_{ap}$ indicate the oxygen ions between the NN Mn ions along the $c$ direction with parallel and antiparallel spins, respectively. All displacements are in \AA. }
\begin{tabular}{p{38pt}|p{48pt}p{48pt}p{48pt}|p{48pt}p{48pt}p{48pt}|p{48pt}p{48pt}p{48pt}}
\hline
\hline
& \multicolumn{9}{c}{HoMnO$_3$}
\tabularnewline
& \multicolumn{3}{c}{E-AFM} & \multicolumn{3}{c}{H-AFM} & \multicolumn{3}{c}{I-AFM} 
\tabularnewline
\hline
& \centering{$\Delta x$} & \centering{$\Delta y$} & \centering{$\Delta z$} &\centering{$\Delta x$} & \centering{$\Delta y$} & \centering{$\Delta z$} & \centering{$\Delta x$} & \centering{$\Delta y$} & \centering{$\Delta z$} 
\tabularnewline
\hline
\centering{Mn} & \centering{-0.1187} & \centering{0} & \centering{0} &\centering{0} & \centering{0} & \centering{-0.0096} & \centering{-0.0062} & \centering{0} & \centering{0} 
\tabularnewline
\centering{O$_{p}^{ab}$} & \centering{0.1849} & \centering{0} & \centering{0} &\centering{0} & \centering{0} & \centering{-0.0028} & \centering{-0.0027} & \centering{0} & \centering{0} 
\tabularnewline
\centering{O$_{ap}^{ab}$} & \centering{0.1287} & \centering{0} & \centering{0} &\centering{0} & \centering{0} & \centering{0.0209} & \centering{0.0049} & \centering{0} & \centering{0} 
\tabularnewline
\centering{O$_{p}^{c}$} & \centering{-} & \centering{-} & \centering{-} &\centering{0} & \centering{0} & \centering{-0.0438} & \centering{-0.0071} & \centering{0} & \centering{0} 
\tabularnewline
\centering{O$_{ap}^{c}$} & \centering{-0.1367} & \centering{0} & \centering{0} &\centering{0} & \centering{0} & \centering{0.0408} & \centering{0.0143} & \centering{0} & \centering{0} 
\tabularnewline
\hline
\hline
& \multicolumn{9}{c}{ErMnO$_3$}
\tabularnewline
& \multicolumn{3}{c}{E-AFM} & \multicolumn{3}{c}{H-AFM} & \multicolumn{3}{c}{I-AFM} 
\tabularnewline
\hline
& \centering{$\Delta x$} & \centering{$\Delta y$} & \centering{$\Delta z$} &\centering{$\Delta x$} & \centering{$\Delta y$} & \centering{$\Delta z$} & \centering{$\Delta x$} & \centering{$\Delta y$} & \centering{$\Delta z$} 
\tabularnewline
\hline
\centering{Mn} & \centering{-0.1165} & \centering{0} & \centering{0} &\centering{0} & \centering{0} & \centering{-0.0093} & \centering{-0.0076} & \centering{0} & \centering{0} 
\tabularnewline
\centering{O$_{p}^{ab}$} & \centering{0.1847} & \centering{0} & \centering{0} &\centering{0} & \centering{0} & \centering{-0.0021} & \centering{-0.0041} & \centering{0} & \centering{0} 
\tabularnewline
\centering{O$_{ap}^{ab}$} & \centering{0.1246} & \centering{0} & \centering{0} &\centering{0} & \centering{0} & \centering{0.0202} & \centering{0.0043} & \centering{0} & \centering{0} 
\tabularnewline
\centering{O$_{p}^{c}$} & \centering{-} & \centering{-} & \centering{-} &\centering{0} & \centering{0} & \centering{-0.0439} & \centering{-0.0032} & \centering{0} & \centering{0} 
\tabularnewline
\centering{O$_{ap}^{c}$} & \centering{-0.1354} & \centering{0} & \centering{0} &\centering{0} & \centering{0} & \centering{0.0411} & \centering{0.0141} & \centering{0} & \centering{0} 
\tabularnewline
\hline
\hline
\end{tabular}
\label{tab:displacements}
\end{table*} 
Next we calculate the electric polarizations $P$ which are induced in o-HoMnO$_3$ by the magnetic orders observed in the MC simulations described in the previous section. We start by considering the commensurate magnetic orders (E-AFM, H-AFM and I-AFM). 
We perform Berry phase calculations using the 1$\times$2$\times$2 supercells in which the ionic positions were optimized imposing E-AFM, H-AFM and I-AFM orders (the supercell in which the positions were relaxed with A-AFM order is used as a reference high symmetry structure). The resulting $P$ values are summarized in Table \ref{table_DFT_en}. One can see that E-AFM order gives the largest electric polarization (4.09 $\mu C$/cm$^2$) among these magnetic phases and $P$ is aligned along the $a$ direction in agreement with Refs. \onlinecite{yamauchi2008homno3} and \onlinecite{picozzi2007homno3}. In turn, H-AFM and I-AFM orders induce $P$ values which are at least one order of magnitude smaller than that of the E-AFM order. Moreover, the $P$ arising from the H-AFM order is aligned along the $c$ direction. To clarify the origin of the differences in $P$ for E-AFM, H-AFM and I-AFM orders, we analyze the displacements of Mn and O ions due to the presence of these orders from their positions in the centrosymmetric structure obtained with A-AFM order. The contributions to $P$ due to the displacements of Mn and O ions within the $ab$ and $ac$ planes are obtained using the point charge model (with ionic charges Mn: (3+) and O: (2-)) and presented in Fig.\ \ref{fig:polarizations}. Exact magnitudes of the displacements are summarized in Table \ref{tab:displacements}.
One can see that in the case of E-AFM order the major contribution to $P$ originates from the displacements of the Mn and O ions within the $ab$ planes. This displacement pattern was explained in Ref.\ \onlinecite{yamauchi2008homno3} in terms of asymmetric electron hopping between the $e_g$ orbitals of Mn ions. Namely, hopping of the $e_g$ electrons occurs only between Mn ions with parallel spins within the $ab$ planes and is forbidden between Mn ions with antiparallel spins. The ions shift such as to enhance this hopping by increasing the Mn-O-Mn bond angles between the corresponding Mn ions \cite{yamauchi2008homno3}. The ionic displacements are the same in the $ab$ planes with different $c$ values and, therefore, they reinforce each other and result in strong $P$ aligned along the $a$ axis. For the H-AFM and I-AFM orders the pairs of Mn ions with parallel and antiparallel spins alternate not only within the $ab$ planes, but also along the $c$ direction (in the E-AFM order the spins on the NN Mn ions along the $c$ axis are always antiferromagnetically oriented). The dominating contributions to the superexchange interactions between the Mn spins along the $c$ axis are due to electron hopping between the $t_{2g}$ orbitals (through the $p_\pi$ states of O anions), which occurs only between $t_{2g}$ states with antiparallel spins. Because of the geometry of the participating orbitals, the hopping is almost independent of the Mn-O-Mn bond angles and is defined rather by the Mn-O distances. Therefore, to enhance this hopping, the Mn and O ions move in the direction that reduces the distances between the Mn ions with antiparallel spins.  The overall ionic shifts result in the maximal energy gain from both in-plane and interplane superexchanges. From Fig.\ \ref{fig:polarizations} one can see that for both H-AFM and I-AFM orders the Mn and O ions move within the $ab$ planes similarly to the case of the E-AFM order, but these displacements occur in the opposite directions for the neighboring $ab$ planes (for example, at $c$=0 and $c$=0.25) and, therefore, the corresponding contributions to the electric polarization almost exactly compensate each other. 

In the H-AFM order the spins on NN Mn ions along the $c$ axis form ferromagnetic and antiferromagnetic stripes alternating along the $b$ direction (see Fig.\ \ref{fig:weird_sW} (a)). $P_c$ originates from the inequivalent displacements of the Mn ions with parallel and antiparallel spins ($a$ and $b$ components of the displacement vectors sum up to zero). This in turn favors small shifts of the in-plane oxygen anions along the $c$ direction giving an additional contribution to $P_c$. The $P$ contributions due to the displacements of the interplane oxygens in FM and AFM stripes, in turn, almost fully cancel each other. In the I-AFM order each Mn ion has two NN Mn ions along the $c$ direction, one with parallel and one with antiparallel spin. The Mn ions move so as to bring the NNs with antiparallel spins closer to each other and to separate those with parallel spins. The $a$ components of the displacement vectors provide $P_a$, while the $b$ and $c$ components cancel each other. The magnitude of $P_a$ is reinforced by displacements of the interplane O anions positioned between Mn ions with antiparallel spins.     
Note that in all cases there is a sizable contribution to the electric polarization arising from the distortion of the electronic density favored by the presence of these magnetic orders, as was shown in detail by Yamauchi \textit{et al.} \cite{yamauchi2008homno3} for the E-AFM order in o-HoMnO$_3$.  

Next we consider the incommensurate w-spiral order and calculate the polarization $\mathbf{P^{AS}}$ due to the inverse DMI using Eq.\ \ref{eq:polar_as}. We find that $\mathbf{P^{AS}}$=(-0.003,0,0.769) per spin suggesting a polarization along the $c$ axis for this magnetic order.
The maximal possible absolute value for the components of the vector $\mathbf{P^{AS}}$ is 2, which would occur for a cycloidal spiral with 90 degree angles between the neighboring spins. A spiral with a modulation vector of $q_b$=0.25, which is close to that of TbMnO$_3$, and spins rotating within the $bc$ plane, produces $\mathbf{P^{AS}}$=(0,0,1.41). Therefore, for our predicted spiral state we expect polarization values smaller than those of typical cycloidal spirals. We also calculate the $P^S_{ab}$ and $P^S_c$ contributions to the electric polarization using Eqs.\ \ref{eq:polar_s} and \ref{eq:polar_s_c}, respectively. In both cases we obtain negligibly small values.

Thus we conclude that the w-spiral order is a plausible magnetic ground state for the o-HoMnO$_3$ single crystal investigated by Lee \textit{et al.} as it is consistent with the magnetic measurements and produces an electric polarization in the direction observed experimentally. In turn, the presence of I-AFM order can provide a possible explanation for the spontaneous polarization $P||a$ with small amplitude, which was observed in a different samples of o-HoMnO$_3$ \cite{lorenz2007homno3_ymno3,feng2010homno3}. In this case, however, one should keep in mind that I-AFM order is not entirely consistent with the magnetic measurements performed so far (it does not give peaks in the magnetic structure factors at either (0,$q$,1) or (0,$q$,0)), therefore more investigations in which both magnetic and ferroelectric orders are investigated for the same sample are required.
\begin{figure}
\begin{centering}
\includegraphics[width=0.41\textwidth]{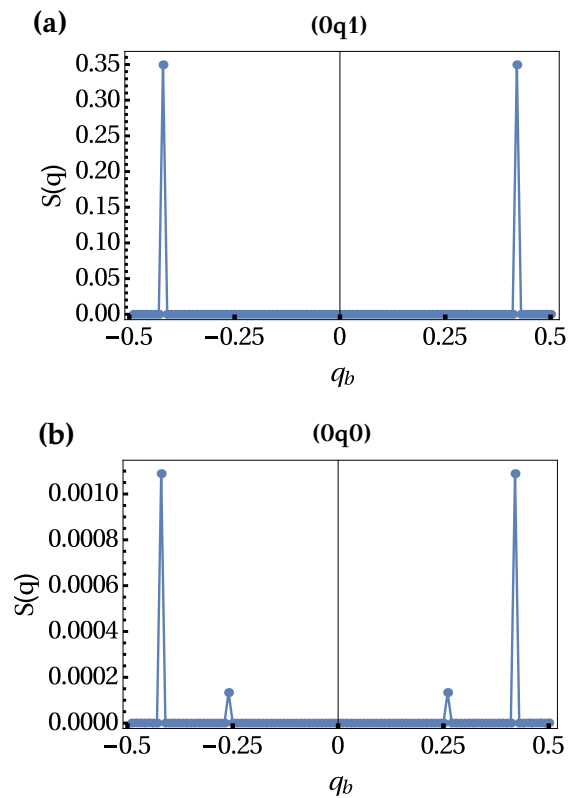}
\caption{\label{fig:wSpiralSFErMnO3} Magnetic structure factors (a) along (0,$q$,1) and (b) along (0,$q$,0) directions in reciprocal space calculated for the  w-spiral ordering observed in MC simulations for o-ErMnO$_3$. $q_b$ is in reciprocal lattice units.}
\end{centering}
\end{figure}

\begin{figure}
\begin{centering}
\includegraphics[width=0.41\textwidth]{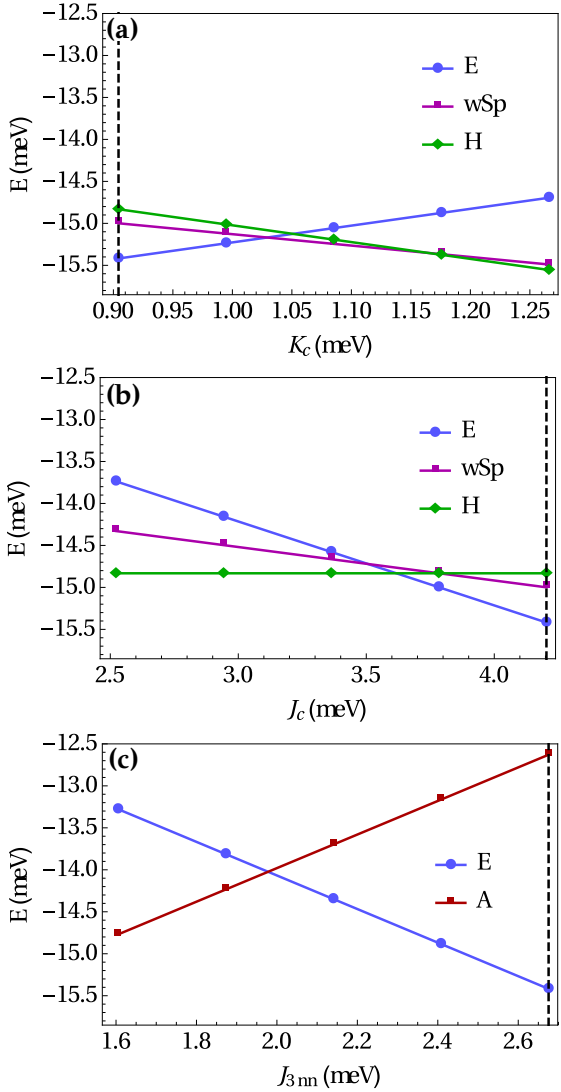}
\caption{\label{fig:en_comp_Er} Exact energies (per spin) of different magnetic orders calculated for o-ErMnO$_3$ as functions of $K_c$, $J_c$ and $J_{3nn}$. (a) shows the energies of E-AFM (E), w-spiral (wSp) and H-AFM (H) orders as functions of $K_c$; (b) shows the energies of the same orders as functions of $J_c$; (c) shows the energies of E-AFM and A-AFM orders as functions of $J_{3nn}$. Dashed lines indicate the initial values of the corresponding couplings extracted using DFT and presented in Table \ref{tab:allcouplings}.}
\end{centering}
\end{figure}

\subsection{ErMnO$_3$}

In this section we investigate the properties of o-ErMnO$_3$ by repeating the procedure described previously for o-HoMnO$_3$. First we theoretically optimize the volume and ionic positions of the experimental crystal structure reported in Ref.\ \onlinecite{ye2007ermno3}. The experimental and optimized lattice parameters are presented in Table \ref{tab:lattice_parameters}. Then we use this relaxed structure to extract a set of microscopic exchange interactions by mapping the results of DFT calculations onto the Hamiltonian of Eq.\ \ref{fullHam}. The resulting couplings are presented in Table \ref{tab:allcouplings}. One can see that these couplings are very similar to those obtained for o-HoMnO$_3$ and changing R from Ho to Er (which results in increased GFO distortion and reduced Mn-O-Mn bond angles within the $ab$ planes) leads only to a reduction (in absolute value) of the in-plane NN exchange $J_{ab}$ (see Fig.\ \ref{fig:exchanges}). Next, we use these couplings to find the ground state of the model Hamiltonian (Eq.\ \ref{fullHam}) in a series of MC simulations. Similarly to o-HoMnO$_3$, we find that E-AFM order with the spins slightly canted away from the $b$ axis is the ground state for o-ErMnO$_3$, but this state is obtained only when the simulations start from the perfect E-AFM order, which indicates strong competition between different exchange interactions resulting in multiple states with very close energies. We also observe that the w-spiral order becomes a ground state if we use in our MC simulations a set of couplings obtained by solving the system of equations with $N_{eq}\neq32$, which differs strongly from that presented in Table \ref{tab:allcouplings} (see the couplings corresponding to $N_{eq}$=33 in Table \ref{tab:couplings_vs_equations} in Appendix \ref{app:exchanges}). The magnetic structure factors calculated for this state show peaks at (0,$\pm$0.42,1) giving $q_b$=0.42, which is in agreement with the experimentally determined value of $q_b$=0.433 \cite{ye2007ermno3}. Similarly to the case of o-HoMnO$_3$, peaks at (0,$\pm$0.42,0) and (0,$\pm$0.26,0) with much smaller intensities were obtained as well (see Fig.\ \ref{fig:wSpiralSFErMnO3}). We also extracted the polarizations $\mathbf{P^{AS}}$, $P^S_{ab}$ and $P^S_c$ using Eqs.\ \ref{eq:polar_as}, \ref{eq:polar_s} and \ref{eq:polar_s_c}, respectively. We find that the w-spiral order induces $\mathbf{P^{AS}}$=(-0.005,0,-0.773). Therefore, $P$ is expected along the $c$ axis in o-ErMnO$_3$ ($P^S_{ab}$ and $P^S_c$ are negligible) if this magnetic order is stabilized.

Next we verify that the w-spiral state occurs in o-ErMnO$_3$ as a result of the interplane four-spin coupling $K_c$ and check whether different magnetic states can be stabilized by changing the other exchange interactions. For this purpose we run MC simulations in which one of the considered couplings is varied by $\pm30\%$ of the value presented in Table \ref{tab:allcouplings} while all the others are kept equal to those in Table \ref{tab:allcouplings}. 
We find that, in contrast to the case of o-HoMnO$_3$, a reduction of $J_{3nn}$ leads to the establishment of A-AFM order. 
The results of the change of the interplane couplings, however, are in agreement with those of o-HoMnO$_3$: an increase in $K_c$ leads to a transition from E-AFM to w-spiral order with $q_b=0.45$, and a reduction of $J_c$ leads to the establishment of H-AFM order with $q_b$=0.5. Since the energy of H-AFM order is equivalent to that of I-AFM order, we perform the following calculations to check which of these two orders will more likely form in o-ErMnO$_3$: We construct a 1$\times$2$\times$2 supercell (the fully relaxed o-ErMnO$_3$ unit cell is doubled along the $b$ and $c$ directions) and relax the ionic positions within this supercell imposing E-AFM, H-AFM and I-AFM orders. Then we use these relaxed structures to calculate the total energies of the o-ErMnO$_3$ supercell with the corresponding magnetic orders and the electric polarizations which are induced by these orders. The results are presented in Table \ref{table_DFT_en}. We find that E-AFM is the lowest energy state in DFT, while I-AFM is the state with the second lowest energy and is more favorable in this system than H-AFM order (by 0.75 meV per spin). The obtained values and the directions of the electric polarizations induced by E-AFM, H-AFM and I-AFM orders in o-ErMnO$_3$ are very close to those calculated for o-HoMnO$_3$. 
\begin{table*}[htb]
\caption{Positions and intensities of the magnetic peaks given by the E-AFM, H-AFM and I-AFM orders, as well as w-spiral, cycloidal $ab$ and $bc$ spirals and sinusoidal order with $q_b$=0.43. The first number gives the position of the peak, the second its intensity (the maximal intensity is 1), "-" indicates that no peak was obtained for this direction in reciprocal space for the corresponding magnetic order.  }
\begin{tabular}{p{62pt}p{70pt}p{70pt}p{70pt}p{70pt}p{70pt}p{70pt}}
\hline
\hline
\centering{($h$,$k$,$l$)} & \centering{E-AFM} & \centering{H-AFM} &  \centering{I-AFM} &  \centering{w-spiral} &  \centering{CS$_{ab}$, CS$_{bc}$} & \centering{SIN}
\tabularnewline

\multirow{2}{62pt}{\centering{(0,$q$,0)}} & \multirow{2}{70pt}{\centering{-}} & \multirow{2}{70pt}{\centering{0.5, 0.25}} &  \multirow{2}{70pt}{\centering{-}} &  \multirow{2}{70pt}{\centering{{0.43}, 1.1$\cdot$10$^{-3}$ \\ 0.29, 1.5$\cdot$10$^{-4}$}} &  \multirow{2}{70pt}{\centering{-}} &  \multirow{2}{70pt}{\centering{-}}
\tabularnewline
\tabularnewline

\centering{(0,$q$,0.5)} & \centering{-} & \centering{-} &  \centering{0.5, 0.5} &  \centering{-} &  \centering{-} &  \centering{-}
\tabularnewline

\centering{(0,$q$,1)} & \centering{0.5, 0.5} & \centering{0.5, 0.25} &  \centering{-} &  \centering{0.43, 0.35} &  \centering{0.43, 0.5} &  \centering{0.43, 0.25}
\tabularnewline

\centering{(0,$q$,1.5)} & \centering{-} & \centering{-} &  \centering{-} &  \centering{-} &  \centering{-}  &  \centering{-}
\tabularnewline

\multirow{2}{62pt}{\centering{(0,$q$,2)}} & \multirow{2}{70pt}{\centering{-}} & \multirow{2}{70pt}{\centering{0.5, 0.25}} &  \multirow{2}{70pt}{\centering{-}} &  \multirow{2}{70pt}{\centering{{0.43}, 1.1$\cdot$10$^{-3}$ \\ 0.29, 1.5$\cdot$10$^{-4}$}} &  \multirow{2}{70pt}{\centering{-}} &  \multirow{2}{70pt}{\centering{-}}
\tabularnewline
\tabularnewline
\centering{(0,1+$q$,0)} & \centering{-} & \centering{0.5, 0.25} &  \centering{-} &  \centering{0.43, 0.15} &  \centering{-}  &  \centering{-}
\tabularnewline

\centering{(0,1+$q$,0.5)} & \centering{-} & \centering{-} &  \centering{-} &  \centering{-} &  \centering{-} &  \centering{-}
\tabularnewline

\centering{(0,1+$q$,1)} & \centering{0.5, 0.5} & \centering{0.5, 0.25} &  \centering{-} &  \centering{0.43, 1.1$\cdot$10$^{-3}$} &  \centering{-}  &  \centering{-}
\tabularnewline

\centering{(0,1+$q$,1.5)} & \centering{-} & \centering{-} &  \centering{0.5, 0.5} &  \centering{-} &  \centering{-}  &  \centering{-}
\tabularnewline

\centering{(0,1+$q$,2)} & \centering{-} & \centering{0.5, 0.25} &  \centering{-} &  \centering{0.43, 0.15} &  \centering{-}  &  \centering{-}
\tabularnewline
\hline
\hline
\end{tabular}
\label{tab:peaks}
\end{table*}

\begin{table*}
\caption{Directions and relative magnitudes of the electric polarizations induced by the E-AFM, H-AFM and I-AFM magnetic orders, as well as the w-spiral, cycloidal $ab$ and $bc$ spirals and sinusoidal order with $q_b$=0.43. }
\begin{tabular}{p{58pt}p{60pt}p{60pt}p{60pt}p{60pt}p{60pt}p{60pt}p{60pt}p{60pt}}
\hline
\hline
\centering{$P$} & \centering{E-AFM} & \centering{H-AFM} &  \centering{I-AFM} &  \centering{w-spiral} &  \centering{CS$_{ab}$} &\centering{CS$_{bc}$} & \centering{SIN}
\tabularnewline

\centering{Direction} & \centering{||$a$} & \centering{||$c$} &  \centering{||$a$} &  \centering{||$c$} &  \centering{||$a$} &  \centering{||$c$} & \centering{-}
\tabularnewline

\centering{Amplitude} & \centering{Large} & \centering{Medium} &  \centering{Medium} &  \centering{Small} &  \centering{Small} &  \centering{Small} & \centering{0}
\tabularnewline

\hline
\hline
\end{tabular}
\label{tab:polarizations}
\end{table*}
Next, we calculate the exact energies of the A-AFM, E-AFM, H-AFM (I-AFM),  CS$_{bc}$ and CS$_{ab}$ and SIN states with $q_b$=0.43 and w-spiral with $q_b$=0.42 using Eq.\ \ref{fullHam} and the couplings from Table \ref{tab:allcouplings}. One can see that with this set of the exchange parameters, E-AFM is the lowest energy state (confirming the result of our MC simulations) and the w-spiral state is the second lowest in energy (differing from the E-AFM order by 0.42 meV per spin). Then we calculate the exact energies of E-AFM, H-AFM (I-AFM) and w-spiral orders as a function of $K_c$ ($K_c$ is varied in the range ($K_c^0$,$K_c^0$+0.4$K_c^0$) with a step of 0.1$K_c$) and $J_c$ (the values of $J_c$ are considered in the interval of ($J_c^0$,$J_c^0-0.4J_c^0$) with a step of -0.1$J_c$). The results are shown in Fig.\ \ref{fig:en_comp_Er}. One can see that an increase in $K_c$ by approximately 13\% (which is just 0.12 meV) from the value of $K_c^0$ presented in Table \ref{tab:allcouplings} favors the establishment of w-spiral order in o-ErMnO$_3$. When $K_c$ is increased by more than 30\% (0.27 meV), the H-AFM order becomes the lowest energy state. Reduction of $J_c$ by $\approx
$14\% (0.59 meV) leads to the stabilization of the H-AFM (I-AFM) order. Finally, by reducing $J_{3nn}$ by more than 25\% (0.67 meV), the A-AFM order can be favored over the E-AFM state.

Thus we demonstrate that in o-ErMnO$_3$, similarly to o-HoMnO$_3$, the strong interplane four-spin ring exchange $K_c$ may lead to the establishment of exotic magnetic orders such as the w-spiral, H-AFM or I-AFM. In general, these three orders, as well as E-AFM order, are close in energy  within the framework of the model of Eq.\ \ref{fullHam} and we assume that any of these states can be stabilized in the real materials depending on the synthesis conditions and quality of the investigated samples. The presence of w-spiral order can explain the result of the neutron diffraction measurements performed for o-ErMnO$_3$ by Ye \textit{et al.} \cite{ye2007ermno3}, in which an IC magnetic order with $q_b$=0.433 was observed. Since the electric polarization induced by w-spiral order is relatively small and appears along the $c$ direction (not in the $a$ direction as usually expected for o-$R$MnO$_3$ with small $R$), this can be a possible explanation for why the electric polarization was not found in the preliminary pyroelectric current measurements performed by this group.  
 
\section{Suggestions for future experiments}
\label{sec:guide}
In this section we summarize the values of the observables expected in systems with the exotic magnetic orders (w-spiral, H-AFM and I-AFM) reported in Sec.\ \ref{sec:results}, together with corresponding values for E-AFM, sinusoidal and cycloidal $ab$ and $bc$ spiral orders (with $q_b=0.43$) to assist in the possible determination of these states in future experimental studies. We perform calculations of the magnetic structure factors along different directions in reciprocal space for all these orders. The w-spiral configuration is adopted from our MC simulations for o-HoMnO$_3$ and all other spin configurations are constructed using the corresponding values of $q_b$ and the relative phases between the spins within the magnetic unit cells. The positions of the peaks and their intensities are presented in Table \ref{tab:peaks} (note that the maximal intensity is 1). In Table \ref{tab:polarizations} we present the directions and relative magnitudes of the electric polarization induced by these orders.

\section{Summary and conclusions}
\label{sec:conclusions}
In summary, we investigated the magnetic and ferroelectric properties of o-HoMnO$_3$ and o-ErMnO$_3$ using \textit{ab initio} calculations and Monte Carlo simulations. The magnetism in these compounds was treated in terms of a model Hamiltonian (Eq.\ \ref{fullHam}), which includes the Heisenberg, biquadratic and four-spin ring exchanges as well as anisotropic terms (DMI and SIA). First, we extracted all the considered microscopic exchange interactions by mapping the results of DFT calculations onto this model Hamiltonian. We found that almost all the coupling constants are relatively large which may result in strong competition between them and lead to magnetic frustration. Then we performed a series of MC simulations using the obtained exchange couplings and found that the magnetic ground state in both systems is the E-AFM order with the spins slightly canted away from the $b$ axis. However, we also observed that small variations of the exchange interactions (within the uncertainty of the method which we used to calculate them) may stabilize other magnetic states such as A-AFM, cycloidal spiral, w-spiral, H-AFM and I-AFM orders). We assume that small differences in the lattice parameters of the experimentally investigated samples (due to different synthesis conditions or the presence of defects) may be enough to provide such a variation of the exchange interactions and may explain the contradictory magnetic measurements. 
\begin{table*}[htb]
\caption{Heisenberg ($J_c$, $J_{ab}$, $J_a$, $J_{diag}$, $J_b$, $J_{3nn}$) and four-spin ring ($K_{ab}$ and $K_c$) exchanges (in meV) in bulk o-HoMnO$_3$ and o-ErMnO$_3$ calculated using the DFT energies of different numbers (16, 20, 24, 28, 32, 33) of inequivalent collinear magnetic configurations.}
\begin{tabular}{p{24pt}|p{26pt}p{26pt}p{26pt}p{26pt}p{26pt}p{26pt}p{26pt}p{26pt}|p{26pt}p{26pt}p{26pt}p{26pt}p{26pt}p{26pt}p{26pt}p{26pt}}
\hline
\hline
& \multicolumn{8}{c}{HoMnO$_3$} & \multicolumn{8}{c}{ErMnO$_3$}
\tabularnewline
\centering{$N$} & \centering{$J_c$} & \centering{$J_{ab}$} & \centering{$J_a$} &\centering{$J_{diag}$} & \centering{$J_b$} & \centering{$J_{3nn}$} & \centering{$K_{ab}$} & \centering{$K_c$} & \centering{$J_c$} & \centering{$J_{ab}$} & \centering{$J_a$} &\centering{$J_{diag}$} & \centering{$J_b$} & \centering{$J_{3nn}$} & \centering{$K_{ab}$} & \centering{$K_c$}
\tabularnewline
\hline 
\centering{16} & \centering{4.33} & \centering{-4.45} & \centering{-0.98} & \centering{0.71} & \centering{1.29} & \centering{2.58} &\centering{0.07} & \centering{0.95} & \centering{4.31} & \centering{-3.78} & \centering{-1.04} &\centering{0.71} & \centering{1.29} & \centering{2.57} & \centering{0.14} & \centering{0.99}
\tabularnewline
\centering{20} & \centering{4.48} & \centering{-4.38} & \centering{-1.04} & \centering{0.75} & \centering{1.28} & \centering{2.56} &\centering{0.14} & \centering{1.02} & \centering{4.44} & \centering{-3.72} & \centering{-1.04} &\centering{0.74} & \centering{1.28} & \centering{2.57} & \centering{0.16} & \centering{1.02}
\tabularnewline
\centering{24} & \centering{4.35} & \centering{-4.43} & \centering{-0.97} & \centering{0.73} & \centering{1.28} & \centering{2.55} &\centering{0.17} & \centering{0.95} & \centering{4.32} & \centering{-3.77} & \centering{-0.96} &\centering{0.72} & \centering{1.28} & \centering{2.56} & \centering{0.19} & \centering{0.95}
\tabularnewline
\centering{28} & \centering{4.30} & \centering{-4.45} & \centering{-0.98} & \centering{0.72} & \centering{0.96} & \centering{2.70} &\centering{0.19} & \centering{0.93} & \centering{4.11} & \centering{-3.86} & \centering{-0.99} &\centering{0.67} & \centering{0.86} & \centering{2.77} & \centering{0.18} & \centering{0.87} 
\tabularnewline
\centering{32} & \centering{4.23} & \centering{-4.49} & \centering{-1.02} & \centering{0.70} & \centering{0.87} & \centering{2.69} &\centering{0.29} & \centering{0.90} & \centering{4.20} & \centering{-3.81} & \centering{-0.99} &\centering{0.69} & \centering{0.95} & \centering{2.68} & \centering{0.28} & \centering{0.90}
\tabularnewline
\centering{33} & \centering{4.98} & \centering{-4.13} & \centering{-1.01} & \centering{0.88} & \centering{0.88} & \centering{2.69} &\centering{0.28} & \centering{1.10} & \centering{4.94} & \centering{-3.46} & \centering{-0.99} &\centering{0.87} & \centering{0.96} & \centering{2.68} & \centering{0.27} & \centering{1.10}
\tabularnewline

\hline
\hline
\end{tabular}
\label{tab:couplings_vs_equations}
\end{table*} 
The key finding of this work is the existence of three new, low energy magnetic orders -- w-spiral, H-AFM and I-AFM -- which are favored by strong interplane four-spin ring interactions (previous works treated the evolution of the magnetic phases in the o-$R$MnO$_3$ series in terms of the competition between the exchange interactions within the $ab$ planes). The presence of the w-spiral order can explain the results of neutron diffraction measurements (Refs. \onlinecite{brinks2001homno3,lee2011homno3,ye2007ermno3}) for o-HoMnO$_3$ and o-ErMnO$_3$, in which incommensurate magnetic orders with $q_b$$\approx$0.4 were found. Since the w-spiral order induces a small electric polarization along the $c$ axis, it can also explain the unexpected polarization direction ($P$||$c$) which was observed by Lee \textit{et al.} \cite{lee2011homno3} in pyroelectric current measurements for o-HoMnO$_3$, and the fact that no sizable $P$ was measured in o-ErMnO$_3$. In turn, the I-AFM order can give rise to $P$||$a$ with small amplitude, which is in agreement with the values reported from the experiments. This, however, should be checked by measuring the magnetic and ferroelectric properties for the same samples of o-HoMnO$_3$ or o-ErMnO$_3$, because magnetic peaks corresponding to I-AFM order have not been experimentally reported to date. 

\section{Acknowledgments}
We thank Andrea Scaramucci for providing the Monte Carlo code and for his guidance during the implementation of the required parts of the model Hamiltonian to this code. We also thank Andrea Scaramucci, Claude Ederer, William Y. Windsor, Urs Staub, Saumya Mukherjee, Christof Niedermayer and Christof W. Schneider for the fruitful discussions. 

This work was supported by ERC Advanced Grant program (No.\ 291151), and by ETH Z\"{u}rich. Computational resources were provided by ETH Z\"{u}rich and Swiss National Supercomputing Centre (CSCS), project No. p504. 

\section{Appendix}
\appendix
\section{Exchange interactions}
\label{app:exchanges}

\begin{figure}
\includegraphics[width=0.45\textwidth]{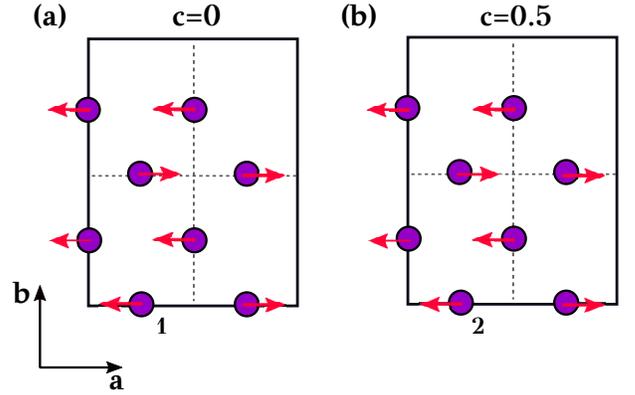}
\caption{\label{fig:structure33} The spin configuration whose DFT energy deviates most from that predicted by the pure Heisenberg Hamiltonian (2$\times$2$\times$1 supercell containing 16 Mn ions. The Mn ions are indicated by the purple spheres. Red arrows show the spins on these ions). It corresponds to C-AFM order with two spins (1 and 2) switched.}
\end{figure}

\begin{figure*}
\includegraphics[width=0.92\textwidth]{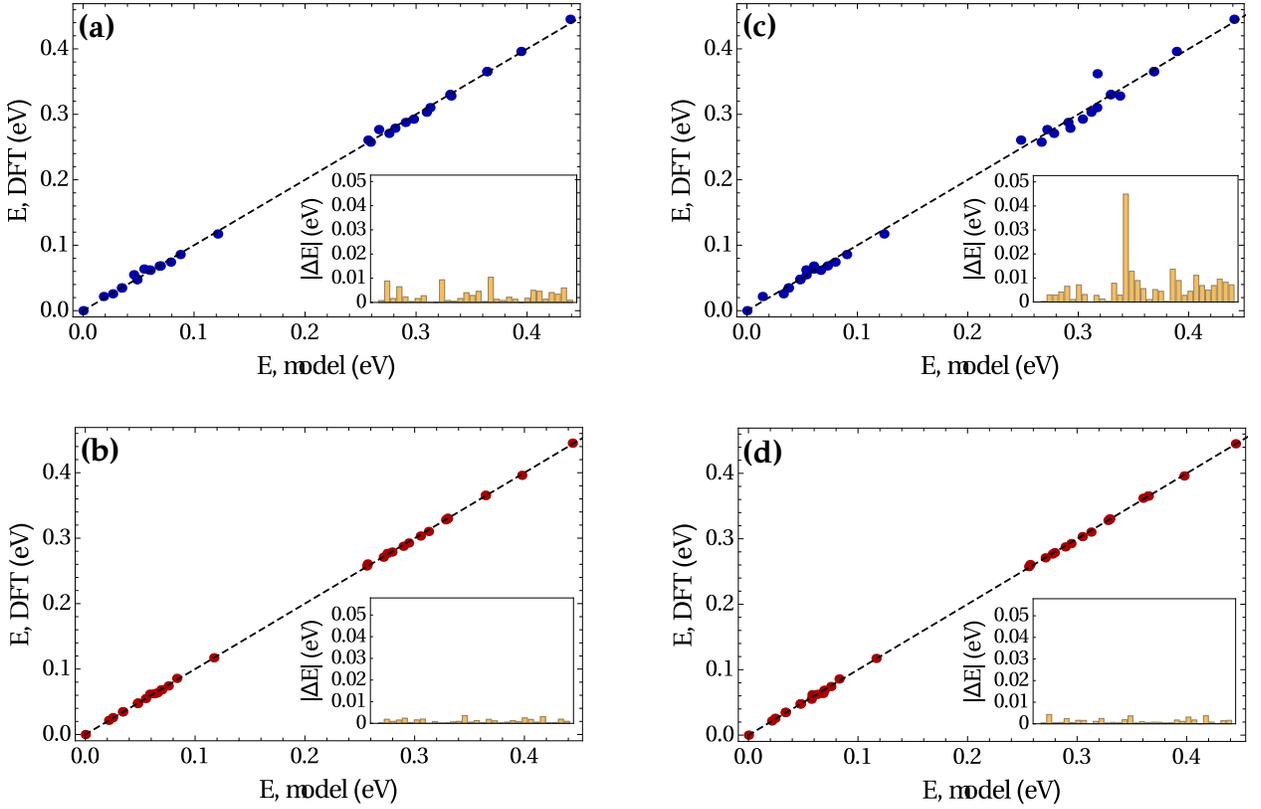}
\caption{\label{fig:exchanges_appendix} Energies of the 80 atom supercell (in eV) of o-HoMnO$_3$ with 32 ((a) and (b)) and 33 ((c) and (d)) inequivalent collinear magnetic orders predicted by different magnetic model Hamiltonians and plotted versus the energies of the corresponding states calculated using DFT (the energy of the E-AFM order is taken as a reference). (a) and (c) show the energies predicted by the pure Heisenberg Hamiltonian, (c) and (d) by the Hamiltonian which includes both Heisenberg and four-spin ring exchanges. The insets on each plot show the differences between the model and DFT energies and each bar corresponds to one of the considered magnetic configurations.}
\end{figure*}

Here we discuss how the Heisenberg and four-spin ring exchange interactions that we calculate for our model Hamiltonian (Eq.\ \ref{fullHam}) are affected by the details of the fitting procedure. As described in Sec.\ \ref{subsec:results_ho_magn}, we extract the Heisenberg and four-spin ring couplings by calculating the DFT energies of many inequivalent collinear magnetic states and constructing an overdetermined system of equations with respect to these couplings. This system is then solved using the least mean square method. The resulting couplings can be affected by the number of equations which are included in the system. In Table \ref{tab:couplings_vs_equations} we present the values of the Heisenberg and four-spin ring couplings which were obtained for o-HoMnO$_3$ and o-ErMnO$_3$ using 16, 20, 24, 28, 32 and 33 equations.

One can see that, for most of the couplings, varying the number of considered equations between 16 and 32 leads to a variation in the obtained values of up to 10\% (the exception is $K_{ab}$, but this coupling is in general very weak). However, the addition of just one equation (33) changes the values of the interplane couplings by up to 25\%. The spin configuration corresponding to equation 33 is a C-AFM order with two spins switched (see Fig.\ \ref{fig:structure33}).

In order to clarify the origin of such a significant change in the values of the interplane couplings we focus on the case of o-HoMnO$_3$ and perform the following analysis. We consider first 32 inequivalent collinear magnetic orders whose energies were calculated for this system using DFT as described in Sec.\ \ref{subsec:results_ho_magn}. We assume that the energies of these states can be described by a pure Heisenberg Hamiltonian (Eq.\ \ref{HeisHam}) which includes only the couplings $J_c$, $J_{ab}$, $J_a$, $J_{diag}$, $J_b$ and $J_{3nn}$ (see Fig.\ \ref{fig:exchanges}). We construct an overdetermined system of equations with respect to these couplings by writing the energies of the 32 considered magnetic configurations within this model Hamiltonian and using the corresponding DFT energies (with respect to the energy of the E-AFM order) as the right-hand sides of these equations. By solving this system of equations we find all the aforementioned Heisenberg couplings. Then we use the extracted couplings to calculate the energies of these 32 magnetic states and plot them versus the energies which were obtained using DFT. The result is shown in Fig.\ \ref{fig:exchanges_appendix} (a) and the inset in this figure shows the difference between the model and DFT energies for each considered state. One can see that, in principle, the DFT energies of many considered configurations cannot be treated accurately within the pure Heisenberg Hamiltonian ($\Delta E$ reaches up to 0.01 eV per supercell with 16 Mn atoms, which corresponds to 0.625 meV per spin). After that we repeat the procedure, but now with state 33 included in the system of equations. The obtained plot of the model energies versus DFT energies is presented in Fig.\ \ref{fig:exchanges_appendix} (c). The point which shows the largest deviation between model and DFT energies corresponds to structure 33 ($\Delta E$ reaches 0.045 eV per supercell with 16 Mn atoms, which gives 2.81 meV per spin). Finally, we add four-spin ring interactions back to the model Hamiltonian and find all the couplings using 32 and 33 equations. Then we calculate again the corresponding model energies and plot them versus the DFT energies. The result is shown in Figs.\ \ref{fig:exchanges_appendix} (b) and (d). Clearly, the addition of the four-spin ring term significantly improves the fitting of the DFT energies onto the model Hamiltonian for all the states, including state 33. Thus we can conclude that the four-spin ring terms are particularly important for describing the energy of this state and this can be a reason why its addition significantly affects the resulting interplane couplings. Since we cannot justify whether this state should be included in the system of equation or excluded from it, we take the uncertainty in the values of the extracted exchange couplings to be $\pm25\%$. 
\begin{figure}
\begin{centering}
\includegraphics[width=0.3\textwidth]{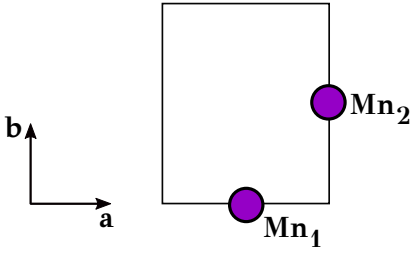}
\caption{\label{fig:MC_unitcell} Two-atom unit cell of o-$R$MnO$_3$ which is used in our MC simulations. Mn ions are shown by the purple spheres.}
\end{centering}
\end{figure}
\section{Magnetic order parameters}
In our MC simulations we consider unit cells containing two Mn atoms with the coordinates Mn$_1$: (0,0.5,0) and Mn$_2$: (0.5,1,0) as shown in Fig.\ \ref{fig:MC_unitcell}. We refer to these as MC unit cells in the following. Note that in this case Mn ions with the coordinates (0,0.5,1) and (0.5,1,1) correspond to the Mn sites with the coordinates (0,0.5,0.5) and (0.5,1,0.5) of the o-$R$MnO$_3$ crystallographic unit cell. We perform simulations with system sizes $N$=$N_a$$\times$$N_b$$\times$$N_c$ MC unit cells, where $N_a$, $N_b$ and $N_c$ are the numbers of MC unit cells along the $a$, $b$ and $c$ directions, respectively. In order to determine the type of ground state Mn spin ordering obtained in the simulations we calculate the following order parameters:

(i) A-AFM order:
\begin{equation}
\bm{\mu}^A=\frac{1}{2N}\sum_{x,y,z=0}^{N_a(N_b, N_c)-1} (-1)^z\left(\mathbf{S}_1^{xyz}+\mathbf{S}_2^{xyz}\right),
\label{eq:OPA}
\end{equation}
where $x$, $y$ and $z$ enumerate the MC unit cells along the $a$, $b$ and $c$ directions, $\mathbf{S}_1^{xyz}$ is the spin of Mn$_1$ in the MC unit cell with certain $x$, $y$ and $z$ and $\mathbf{S}_2^{xyz}$ is the spin of Mn$_2$ in the same MC unit cell. For perfect A-AFM order with the spins aligned along the $b$ axis $\bm{\mu}^A$=(0,$\pm$1,0), the sign is different for the two different orientations of A-AFM domains.

(ii) E-AFM order:
\begin{equation}
\bm{\mu}^E_1=\frac{1}{2N}\sum_{x,y,z=0}^{N_a(N_b, N_c)-1} (-1)^{y+z}\left(\mathbf{S}_1^{xyz}+\mathbf{S}_2^{xyz}\right),
\label{eq:OPE1}
\end{equation}

\begin{equation}
\bm{\mu}^E_2=\frac{1}{2N}\sum_{x,y,z=0}^{N_a(N_b, N_c)-1} (-1)^{y+z}\left(\mathbf{S}_1^{xyz}-\mathbf{S}_2^{xyz}\right).
\label{eq:OPE2}
\end{equation}

$\pm\bm{\mu}^E_1$ and $\pm\bm{\mu}^E_2$ account for the four types of E-AFM domains.

(iii) H-AFM order.

The general expression for the H-AFM order parameters is easier to write in terms of the supercell containing four Mn ions (our MC unit cell is doubled along the $c$ direction): Mn$_1$: (0,0.5,0), Mn$_2$: (0.5,1,0), Mn$_3$: (0,0.5,1) and Mn$_4$: (0.5,1,1). Note, that this MC supercell is equivalent to the o-$R$MnO$_3$ crystallographic unit cell (only Mn ions are considered). 
\begin{eqnarray}
\bm{\mu}^{H}_1=\frac{1}{4N^s}\sum_{x,y,z=0}^{N_a^s(N_b^s, N_c^s)-1} (-1)^{y}\left(\mathbf{S}_1^{xyz} \right. \nonumber \\ + \left.\mathbf{S}_2^{xyz} + \mathbf{S}_3^{xyz}-\mathbf{S}_4^{xyz}\right),
\label{eq:OPsW1}
\end{eqnarray}
where $x$, $y$ and $z$ enumerate the supercells along the $a$, $b$ and $c$ directions; $N_a^s=N_a$, $N_b^s=N_b$, $N_c^s=N_c/2$ are the number of supercells along each direction and $N^s$=$N_a^s$$\times$$N_b^s$$\times$$N_c^s$ gives the total number of supercells. 
For other types of H-AFM domains:
\begin{eqnarray}
\bm{\mu}^{H}_2=\frac{1}{4N^s}\sum_{x,y,z=0}^{N_a^s(N_b^s, N_c^s)-1} (-1)^{y}\left(\mathbf{S}_1^{xyz} \right. \nonumber \\ + \left.\mathbf{S}_2^{xyz} - \mathbf{S}_3^{xyz}+\mathbf{S}_4^{xyz}\right),
\label{eq:OPsW2}
\end{eqnarray}

\begin{eqnarray}
\bm{\mu}^{H}_3=\frac{1}{4N^s}\sum_{x,y,z=0}^{N_a^s(N_b^s, N_c^s)-1} (-1)^{y}\left(\mathbf{S}_1^{xyz} \right. \nonumber \\ - \left.\mathbf{S}_2^{xyz} + \mathbf{S}_3^{xyz}+\mathbf{S}_4^{xyz}\right),
\label{eq:OPsW3}
\end{eqnarray}

\begin{eqnarray}
\bm{\mu}^{H}_4=\frac{1}{4N^s}\sum_{x,y,z=0}^{N_a^s(N_b^s, N_c^s)-1} (-1)^{y}\left(-\mathbf{S}_1^{xyz} \right. \nonumber \\ + \left.\mathbf{S}_2^{xyz} + \mathbf{S}_3^{xyz}+\mathbf{S}_4^{xyz}\right).
\label{eq:OPsW4}
\end{eqnarray}
In total there are 8 types of H-AFM domains ($\pm\bm{\mu}^{sW}_i$, $i$=1,...,4).

(iv) I-AFM order.
Similarly to the H-AFM case we write the expression for the order parameters for I-AFM order using the 4-atom supercell. The equations have the same form as for the H-AFM order with the only difference being $(-1)^y$ in each equation is replaced with $(-1)^{y+z}$.  

\section{Magnetic structure factors}
In order to identify the modulation vectors for the magnetic states which we obtain in our MC simulations, we calculate the absolute values of the magnetic structure factors for Mn moments along different directions in reciprocal space such as:
\begin{equation}
\label{eq:struct_fact1}
S(q)=\mathbf{s}^*(\mathbf{q})\cdot\mathbf{s}(\mathbf{q}),
\end{equation}
where
\begin{equation}
\label{eq:struct_fact2}
\mathbf{s}(\mathbf{q})=\frac{1}{2N}\sum_{i=1}^{2N}\mathbf{S}_ie^{2\pi i\mathbf{R}_i\mathbf{q}}.
\end{equation}
$i$ enumerates the Mn ions in the considered system, $N$ is the number of MC unit cells (each containing 2 Mn ions), $\mathbf{S}_i$ is the Mn spin on site $i$, $\mathbf{R}_i$ is a position of site $i$; in most cases we consider $\mathbf{q}$=(0,$q_y$,0) and (0,$q_y$,1) and $q_y$ takes the values from [-0.5,0.5] with a step of 1/$N_b$, where $N_b$ is the number of unit cells along the $b$ direction. 

\smallskip

\section{Electric polarizations}
The contribution to the electric polarization $\mathbf{P^{AS}}$ due to the inverse Dzyaloshinskii-Moriya interaction is calculated using the following formula:
\begin{eqnarray}
\mathbf{P^{AS}}=\frac{1}{4N} \mathbf{e}_b \times \sum_{i=1}^{2N} \left[\mathbf{S}_i \times (\mathbf{S}_{i+\frac{a}{2}+\frac{b}{2}} \right. \nonumber \\ + \left.\mathbf{S}_{i-\frac{a}{2}+\frac{b}{2}}-\mathbf{S}_{i+\frac{a}{2}-\frac{b}{2}}-\mathbf{S}_{i+\frac{a}{2}-\frac{b}{2}}) \right],
\label{eq:polar_as}
\end{eqnarray}
where $\mathbf{e}_b$ is a unit vector along the $b$ direction,  $\mathbf{S}_i$ is the spin on site $i$, the $S_{i\pm\frac{a}{2}\pm\frac{b}{2}}$ are the spins on the NN Mn sites within the $ab$ planes with respect to $\mathbf{S}_i$ and $N$ is the number of MC unit cells in the considered system. 

The $P^S$ contributions (of spin origin) are obtained from:
\begin{eqnarray}
P^{S}_{ab}=\frac{1}{4N} \sum_{i=1}^{2N} \left(\mathbf{S}_i \cdot (\mathbf{S}_{i+\frac{a}{2}+\frac{b}{2}} \right. \nonumber \\ + \left.\mathbf{S}_{i-\frac{a}{2}+\frac{b}{2}}-\mathbf{S}_{i+\frac{a}{2}-\frac{b}{2}}-\mathbf{S}_{i+\frac{a}{2}-\frac{b}{2}}) \right)
\label{eq:polar_s}
\end{eqnarray}
within the $ab$ planes, and
\begin{eqnarray}
P^{S}_{c}=\frac{1}{4N} \sum_{i=1}^{2N} \left(\mathbf{S}_i \cdot (\mathbf{S}_{i+c} - \mathbf{S}_{i-c}) \right)
\label{eq:polar_s_c}
\end{eqnarray}
along the $c$ axis. Since our MC unit cells contain only 2 atoms, $c$ in Eq.\ \ref{eq:polar_s_c} corresponds to $c/2$ of the crystallographic unit cell of o-$R$MnO$_3$. 
Note, that $P^S_{ab}$ and $P^S_c$ as well as $\mathbf{P^{AS}}$ do not give estimates for the magnitudes of the electric polarizations induced by the corresponding mechanisms. This is because the ionic displacements are not considered, there are no coefficients accounting for the difference between NN Mn sites with FM and AFM oriented spins for the collinear spin orders, and the fact that $P$ induced by symmetric exchange striction is larger than that arising due to weak DMI is not taken into account.

\bibliographystyle{apsrev}
\bibliography{main}
\end{document}